\documentclass{AIAA}
\usepackage{amsmath}
\usepackage{amssymb}
\usepackage{subfigure}
\usepackage{tabularx}
\usepackage{url}

\DeclareMathOperator{\round}{round}
\DeclareMathOperator{\E}{E}

\begin{document}

\title{Numerical Error in Interplanetary Orbit Determination Software}

\author{Marco Zannoni\footnote{Ph.D. student, Department of Mechanical and Aerospace Engineering, via Fontanelle 40, \url{m.zannoni@unibo.it.}} and Paolo Tortora\footnote{Associate Professor, Department of Mechanical and Aerospace Engineering, via Fontanelle 40, \url{paolo.tortora@unibo.it}, AIAA Senior Member}}
\affiliation{University of Bologna, 47121 Forl\`i, Italy}

\begin{abstract}
The core of every orbit determination process is the comparison between the measured observables and their predicted values, computed using the adopted mathematical models, and the minimization, in a least square sense, of their differences, known as residuals. In interplanetary orbit determination, Doppler observables, obtained by measuring the average frequency shift of the received carrier signal over a certain count time, are compared against their predicted values, usually computed by differencing two round-trip light-times. This formulation is known to be sensitive to round-off errors, caused by the use of finite arithmetic in the computation, giving rise to an additional noise in the residuals, called numerical noise, that degrades the accuracy of the orbit determination solution. This paper presents a mathematical model for the expected numerical errors in two-way and three-way Doppler observables, computed using the differenced light-time formulation. The model was validated by comparing its prediction to the actual noise in the computed observables, obtained by NASA/Jet Propulsion Laboratory's Orbit Determination Program. The model proved to be accurate within $3 \times 10^{-3} \,\text{mm/s}$ at $60 \,\text{s}$ integration time. Then it was applied to the case studies of Cassini's and Juno's nominal trajectories, proving that numerical errors can assume values up to $6 \times 10^{-2} \,\text{mm/s}$ at $60 \,\text{s}$ integration time, and consequently that they are an important noise source in the Doppler-based orbit determination processes. Three alternative strategies are proposed and discussed in the paper to mitigate the effects of numerical noise.
\end{abstract}

\maketitle

\section*{Nomenclature}
\noindent\begin{tabularx}{\textwidth}{@{}lcX@{}}
$b_i$           &=&     digits of the binary floating point representation of $f$ \\
$c$             &=&     speed of light in vacuum (km/s) \\
$f$             &=&     absolute value of fractional part of the mantissa in the binary floating point representation \\
$f_T$           &=&     transmitting frequency at the transmitting ground station (Hz) \\
$F_2$           &=&     two-way Doppler observable (Hz) \\
$F_3$           &=&     three-way Doppler observable (Hz) \\
$M_2$           &=&     spacecraft transponder turnaround ratio, which is the ratio of the transmitted frequency to the received frequency at the spacecraft \\
$p$             &=&     exponent (base 2) of the most significant digit in the binary floating point representation \\
$q$             &=&     maximum value represented by the least significant bit in the binary floating point representation \\
$\mathbf{r_1}$  &=&     position vector of the transmitting ground station at $t_1$, with respect to the Solar System barycenter (km) \\
$\mathbf{r_2}$  &=&     position vector of the spacecraft at $t_2$, with respect to the Solar System barycenter (km) \\
$\mathbf{r_3}$  &=&     position vector of the receiving ground station at $t_3$, with respect to the Solar System barycenter (km) \\
$r_{12}$    &=&     Newtonian distance between the transmitting ground station at $t_1$, and the spacecraft at $t_2$ (km) \\
$r_{23}$    &=&     Newtonian distance between the spacecraft at $t_2$, and the receiving ground station at $t_3$ (km) \\
$\dot{r}_{12}$  &=&     Newtonian range-rate between the transmitting ground station, at the transmission time, and the spacecraft, at the reflection time (km/s) \\
$\dot{r}_{23}$  &=&     Newtonian range-rate between the spacecraft, at the reflection time, and the receiving ground station, at the reception time (km/s)
\end{tabularx}

\noindent\begin{tabularx}{\textwidth}{@{}lcX@{}}
$\mathbf{r^i_j}$        &=&     position vector of the body $j$ with respect to the body $i$ (km) \\
$\mathbf{\dot{r}^i_j}$  &=&     relative velocity vector of the body $i$ with respect to the body $j$ (km) \\
$r_s$           &=&     distance of the ground station from the Earth spin axis (km) \\
$R_{\Delta \overline{x}}(\tau)$     &=&     autocorrelation function of $\Delta \overline{x}$, with lag $\tau$ \\
$R_{\Delta \overline{x}}(t,\tau)$   &=&     autocorrelation function of $\Delta \overline{x}$, at time $t$ and with lag $\tau$ \\
$s$     &=&     exponent that defines the sign in the binary floating point representation \\
$t$     &=&     number of binary digits used to represent $f$ \\
$t_1$           &=&     transmission time at the transmitting ground station, in Ephemeris Time (s) \\
$t_2$           &=&     reflection time at the spacecraft, in Ephemeris Time (s) \\
$t_3$           &=&     reception time at the receiving ground station, in Ephemeris Time (s) \\
$t_1(ST)_T$     &=&     transmission time at the transmitting electronics, in Station Time (s) \\
$t_3(ST)_R$     &=&     reception time at the receiving electronics, in Station Time (s) \\
$t_p$   &=&     time elapsed since the start of the current tracking pass (s) \\
$T_c$   &=&     count time interval of the Doppler observables (s) \\
$TT$    &=&     time-tag of a Doppler observable (s) \\
$x_0$   &=&     exact value to be encoded using the binary floating point representation \\
$x$     &=&     binary floating point representation of $x_0$ \\
$\alpha$        &=&     spacecraft right ascension (rad) \\
$\alpha_0$    &=&     right ascension of the prime meridian at the adopted epoch (rad) \\
$\delta$        &=&     spacecraft declination (rad) \\
$\Delta \overline{x}$               &=&     absolute rounding error in the binary floating point representation of $x_0$ \\
$\Delta \tilde{x}$                  &=&     relative rounding error in the binary floating point representation of $x_0$ \\
$\Delta \overline{x}_{\text{max}}$  &=&     maximum absolute value of $\Delta \overline{x}$ \\
$\varepsilon$   &=&     maximum absolute value of $\Delta \overline{x}$, also called machine epsilon \\
$\lambda_s$     &=&     longitude of the ground station (rad) \\
$\mu_i$         &=&     gravitational parameter of the body $i$ (km$^3$/s$^2$) \\
$\rho$          &=&     precision round-trip light-time
\end{tabularx}

\noindent\begin{tabularx}{\textwidth}{@{}lcX@{}}
$\sigma_y(\tau)$                    &=&     Allan standard deviation with integration time $\tau$ \\
$\sigma_{\Delta \overline{x}}$      &=&     standard deviation of $\Delta \overline{x}$ \\
$\sigma^2_{\Delta \overline{x}}$    &=&     variance of $\Delta \overline{x}$ \\
$\omega_e$      &=&     Earth mean rotation rate (rad/s)
\end{tabularx}

\section{Introduction}

The aim of the orbit determination (OD) process is the estimation of a set of parameters that unambiguously defines the trajectory of a spacecraft. The core of the process is the comparison between measured observables (observed observables) and the corresponding computed values (computed observables), obtained by an OD program using the adopted mathematical models. The output of the OD is the values of the parameters (solve-for parameters) that minimize, usually in a least-square sense, the global difference between the observed observables and the computed observables.
For the tracking of deep space probes the use of Earth-based radiometric techniques, especially based on Doppler measurements, is of fundamental importance.

Disturbances in either the observed or computed observables cause errors in the OD process.
The different sources of noise in the Doppler observed observables were the subject of several detailed studies in the past~\cite{Armstrong:2003, Asmar:2005, Armstrong:2006, Armstrong:2008}. In~\cite{Asmar:2005} a detailed noise budget for Doppler tracking of deep space probes was presented; however recent tracking data acquired from the NASA/ESA/ASI Cassini-Hyugens mission to the Saturn system showed imperfect quantitative agreement between the measured residual noise~\cite{Iess:2012} and the predictions presented in~\cite{Asmar:2005}, thus suggesting that there could be additional effects not considered so far, like errors in the computed Doppler observables.

Errors in the computed radiometric observables are caused by incomplete mathematical models implemented in the OD program and by numerical errors that arise from the use of finite-precision arithmetic. Concerning the model errors, the basis of the two most important interplanetary orbit determination programs, NASA/Jet Propulsion Laboratory's (JPL) Orbit Determination Program (ODP) and ESA/ESOC's Advanced Modular Facility for Interplanetary Navigation (AMFIN), is Moyer's formulation, described in detail in~\cite{Moyer:2000}. This formulation is considered sufficiently accurate, compared to the present level of noise in the radiometric measurements.

Numerical errors are introduced in every real number and computation step because, within a computer, the representation of numbers and the computations have to be performed using a finite number of digits. The resulting errors in the computed observables depend upon the hardware and software representation of numbers, the mathematical formulation of the observables and its implementation in the software. According to Moyer's formulation the two-way (or three-way) Doppler observable is computed as the difference of two round-trip light times between the spacecraft and the tracking station(s), thus it is called differenced-range Doppler (DRD) formulation~\cite{Moyer:1969, Moyer:1971}. AMFIN implements also an older formulation, based upon a truncated Taylor series, called the integrated Doppler (ID) formulation, formerly implemented in the ODP and described in~\cite{Moyer:1971}.
Moreover, both the ODP and AMFIN are compiled to use double-precision floating point arithmetic~\cite{Moyer:2000,AMFIN:ALG}.

An order-of-magnitude estimation of numerical errors in DRD double-precision computed observables is provided in~\cite{Moyer:1971} and~\cite{Moyer:1969}, where it is recommended to use at least 60 bits to represent the significand, also called mantissa, in order to keep a two-way accuracy of $2 \times 10^{-2} \, \text{mm/s}$ at all integration times. This results in an Allan Standard Deviation (ASDEV) of about $6.7 \times 10^{-14}\, \text{s/s}$.
However, the target accuracy has been gradually and continuously improving with time. For example, the current most precise two-way Doppler observations were achieved between the NASA Deep Space Network (DSN) and the Cassini spacecraft, reaching a fractional frequency stability, expressed in terms of ASDEV, of about $3 \times 10^{-15}$ at $1000 \, \text{s}$ integration time~\cite{Armstrong:2003}.
Future interplanetary missions, such as ESA's BepiColombo mission to Mercury and NASA's Juno mission to Jupiter, carry radio science instrumentation which, used in conjunction with highly performing ground antennas, is required to reach a two-way fractional frequency stability of $1.0 \times 10^{-14}$ and $8.2 \times 10^{-15}$ at $1000 \, \text{s}$ integration time, respectively~\cite{Iess:2009, Anderson:2004}, during most of their operational life.
On the other hand, almost all modern computers follow the IEEE-754 standard, which establishes the use of 64 bits for double-precision floating point binary representation, where 52 bits are used for the mantissa~\cite{IEEE-754:2008}.
For these reasons a more detailed study of the numerical noise is necessary to assess its actual impact on the precision of the computed observables.

This paper describes a model of numerical errors in two-way (and three-way) Doppler observables, computed by an OD program that implements Moyer's DRD formulation.
In Sec.~\ref{sec:num_err_theory} the basic concepts of the floating point representation and arithmetic are introduced, together with a statistical model used to evaluate the single contributions to the numerical error, their propagation and accumulation. An analytical model to predict numerical errors in two-way differenced-range Doppler observables is presented in Sec.~\ref{sec:num_err_model}, while the validation of this model is described in Sec.~\ref{sec:model_validation}. In Sec.\ref{sec:num_noise_analysis}, the numerical error prediction model is used to analyze the expected numerical noise for Cassini's and Juno's Doppler tracking data. Then, three different strategies to reduce the numerical noise in the Doppler link are presented in Sec.~\ref{sec:num_noise_reduction}. Finally, Sec.~\ref{sec:conclusions} summarizes the results and presents the conclusions of this work.

\section{Round-Off Errors}\label{sec:num_err_theory}
\subsection{Floating Point Representation}
The most used method for representing real numbers in modern computers is the binary floating point representation, based upon the normalized scientific notation in base $2$. Using this representation an exact value $x_0$ is encoded using a finite number of bits, as:
\begin{equation}
x = (-1)^s (1.b_1 b_2 \ldots b_{t-1} b_t) \times 2^p
\end{equation}
where $s=0$ for positive values, $s=1$ for negative ones. $p$ can be computed from $x_0$ using the following relation:
\begin{equation}\label{eq:p_float}
p(x_0) =
\begin{cases}
    \lfloor \log_2{|x_0|} \rfloor, & \text{if $\quad x_0 \neq 0$} \\
    0, & \text{if $\quad x_0 = 0$}
\end{cases}
\end{equation}
where $\lfloor a \rfloor$ is the floor function of $a$.

$b_t$ is the last represented digit of $x$ and is called Least-Significant Bit ($LSB$). In the representation of $x_0$ the maximum value represented by the $LSB$ is:
\begin{equation}
q(x_0) = 2^{p(x_0)-t}
\end{equation}

There are several rounding algorithms that define how to choose $b_t$. The mostly used algorithm is ``round toward nearest, ties to even'', because it minimizes the rounding error for a given $t$ and does not introduce a systematic drift~\cite{Goldberg:1991}. In what follows, only this algorithm will be considered.

The use of a finite number of digits to represent the mantissa introduces an error. The absolute and relative rounding errors are defined as:
\begin{gather}
\Delta \overline{x}(x_0) \triangleq x -x_0 \\
\Delta \tilde{x}(x_0) \triangleq (x -x_0)/x_0
\end{gather}

The maximum absolute value of the absolute error in the representation of a variable $x_0$ is:
\begin{equation}
\Delta \overline{x}_{\text{max}}(x_0) \triangleq \max{|\Delta \overline{x}(x_0)|} = q(x_0)/2 = 2^{p(x_0)-t-1}
\end{equation}
Hence, the maximum absolute error depends upon the number of digits available for the fractional part of the mantissa $t$ and the order of magnitude $p$ of the binary number. Given $t$, the maximum error is the same for all values in the range $[2^p,2^{p+1})$.

Given two proportional values $x_1$ and $x_2=k\,x_1$, in general $\Delta \overline{x}_{\text{max}}(x_2)$ and $\Delta \overline{x}_{\text{max}}(x_1)$ do not satisfy the same relation of proportionality, because of the presence of the floor function in the definition of $p$ (Eq.~\ref{eq:p_float}).
If and only if the absolute value of the scale factor $k$ is an integer power of the basis of representation ($|k| = 2^{n_k}$, $n_k \in \mathbb{Z}$) the maximum absolute errors are proportional, with scale factor $|k|$:
\begin{equation}
    p(x_2) = \lfloor \log_2{|k x_1|} \rfloor = \lfloor \log_2{|k|} + \log_2|x_1| \rfloor = \lfloor n_k + \log_2|x_1| \rfloor = n_k + \lfloor \log_2|x_1| \rfloor = n_k + p(x_1)
\end{equation}
\begin{equation}
    \Delta \overline{x}_{\text{max}}(x_2) = 2^{p(x_2)-t-1} = 2^{n_k + p(x_1)-t-1} = 2^{n_k}\,2^{p(x_1)-t-1} = |k|\,\Delta \overline{x}_{\text{max}}(x_1)
\end{equation}
However, it can be shown that for a generic scale factor $k$ the ratio $\Delta \overline{x}_{\text{max}}(x_2)/\Delta \overline{x}_{\text{max}}(x_1)$ is still $|k|$, when averaged on a large enough interval.

As an immediate consequence, the maximum absolute error in the floating point representation of a specific physical quantity depends upon the adopted unit.
For example:
\begin{gather}
\text{LSB}(1 \, \text{km}) \simeq 2.22 \times 10^{-16} \, \text{km} = 2.22 \times 10^{-13} \, \text{m} \\
\text{LSB}(1000 \, \text{m}) \simeq 1.14 \times 10^{-13} \, \text{m}
\end{gather}
However, averaging on a large enough interval of values, the two maximum absolute errors are equal, so they can be considered equivalent.

The maximum absolute value of the relative error, also called machine epsilon or $\varepsilon$, is:
\begin{equation}
\varepsilon \triangleq \max{\left| \Delta \tilde{x}(x_0) \right|} = \Delta \overline{x}_{\text{max}}/|x_{\text{min}}| = 2^{-t-1}
\end{equation}
$\varepsilon$ does not depend upon the order of magnitude of the value $x_0$, but only upon the number of digits of the fractional part of the mantissa $t$.

The IEEE-754 standard defines several basic floating-point binary formats that differ by the number of bits used to encode the sign, the exponent and the fractional part of the mantissa. The most important formats are Binary32 (also called single precision), Binary64 (also called double-precision), and Binary128 (also called quadruple-precision). Their characteristics are described in Table~\ref{tab:IEEE_formats}~\cite{IEEE-754:2008}.

\begin{table}
\caption{Characteristics of the most important binary formats defined by IEEE-754}\label{tab:IEEE_formats}
\begin{ruledtabular}
\begin{tabular}{cccccc}
                                & \multicolumn{4}{c}{Number of bits}        &       \\\cline{2-5}
IEEE format                     & Total     & $s$       & $f$       & $p$   & $\varepsilon$ \\\hline
Binary32 (Single-Precision)     & 32        & 1         & 23        & 8     & $5.96 \times 10^{-8}$  \\
Binary64 (Double-precision)     & 64        & 1         & 52        & 11    & $1.11 \times 10^{-16}$ \\
Binary128 (Quadruple-Precision) & 128       & 1         & 112       & 15    & $9.63 \times 10^{-35}$  \\
\end{tabular}
\end{ruledtabular}
\end{table}

\subsection{Statistical Model}\label{subsec:stat_mod}
The rounding error is a form of quantization error, with the $\text{LSB}$ as the quantization step. Hence, it is possible to adopt a statistical model, usually applied to quantization errors, known as uniform quantization error model, or Widrow's model~\cite{Widrow:1961}. Using this description, the round-off error $\Delta \overline{x}(t)$ in the numerical representation of an exact time function $x_0(t)$ is modeled as a white stochastic process, wide-sense stationary, uncorrelated with the represented function $x_0(t)$, and with a Probability Density Function (PDF) uniform between the extreme values $-q(x_0)/2$ and $q(x_0)/2$.

With this model, it is possible to compute the statistical characteristics of the numerical error:
\begin{enumerate}
    \item[1)] Mean value: $\E[\Delta \overline{x}(t)]=0$
    \item[2)] Maximum absolute value: $\Delta \overline{x}_{\text{max}} = q/2$
    \item[3)] Variance: $\sigma_{\Delta \overline{x}}^2 = q^2/12$
\end{enumerate}

The model is valid if the PDF and the Characteristic Function (CF) (the Fourier transform of the PDF) of the input signal $x_0(t)$ satisfy the conditions of the Quantization Theorem (QT), stated in~\cite{Widrow:1961}. In~\cite{Sripad:1977} weaker necessary and sufficient conditions for the QT are provided but, in practice, for real-world input signals, the conditions are very restrictive. For example, a sinusoidal signal does not satisfy the QT, but Widrow's model can still be applied with sufficient accuracy if the amplitude of the signal is several times larger than the quantization step~\cite{Mariano:2006}.

In this work the statistical model was assumed to be valid and this assumption was verified through a detailed analysis of the rounding errors, described in Sec.~\ref{subsec:model_complete}.

\subsection{Operation Errors}\label{subsec:num_err_theory_oper}
When performing an exact operation $f(x_0,y_0)$ using floating point arithmetic, the result is affected by two kind of errors with respect to the theoretical value $z_0$:
\begin{enumerate}
    \item[1)] Propagation error of the rounding errors in the inputs:
    \begin{equation}
        x = x_0 + \Delta \overline{x} ,\qquad y = y_0 + \Delta \overline{y}
    \end{equation}
    If the errors are small enough:
    \begin{equation}
        \tilde{z}_0 = f(x,y) \simeq f(x_0,y_0) + \frac{\partial f}{\partial x}(x_0,y_0) \Delta \overline{x} + \frac{\partial f}{\partial y}(x_0,y_0) \Delta \overline{y}
    \end{equation}
    \item[2)] Additional rounding error: even if the errors in the inputs are zero, the output could be affected by a rounding error. IEEE Standard 754 requires that the result of the basic operations (addition, subtraction, multiplication, division and square root) must be exactly rounded~\cite{IEEE-754:2008}, i.e. the operation shall be performed as if it first produced an intermediate result correct to infinite precision, and then that result was rounded to the finite number of digits in use.
    The rounding on the result $\tilde{z}_0$ is represented by the addition of the random variable $\Delta \overline{z}$:
    \begin{equation}
        z = \tilde{z}_0 + \Delta \overline{z} = f(x,y) + \Delta \overline{z}
    \end{equation}
    In practice, to obtain an exactly rounded result, it is necessary to perform the basic operations using at least 3 additional binary digits~\cite{Goldberg:1991}. For non-basic operations, the error on the results could be larger and depends on how the operations are implemented. The additional rounding error may be zero if the intermediate result of the operation does not need rounding.
\end{enumerate}

The total error in the result is the sum of the propagation error and the additional rounding error:
\begin{equation}
    \Delta z = \frac{\partial f}{\partial x}(x_0,y_0) \Delta \overline{x} + \frac{\partial f}{\partial y}(x_0,y_0) \Delta \overline{y} + \Delta \overline{z}
\end{equation}
Using Widrow's model for numerical errors $\Delta \overline{x}$, $\Delta \overline{y}$, and $\Delta \overline{z}$, the total error $\Delta z$ is the sum of three random variables and its statistical characteristics derive from the characteristics of each round-off error.

\section{Numerical Error Model}\label{sec:num_err_model}

\subsection{Differenced-Range Doppler Formulation}
According to Moyer's DRD formulation, the unramped two-way or three-way Doppler observable is proportional to the mean time variation of the round-trip light-time (i.e. the mean range-rate divided by $c$), during the count time interval centered in the observable's time-tag~\cite{Moyer:2000}:
\begin{equation}\label{eq:doppler}
    F_{2,3}(TT) = M_2 f_T \left[ \rho(TT+T_c/2) -\rho(TT-T_c/2) \right] / T_c
\end{equation}
where $M_2$ is the spacecraft transponder turnaround ratio, which is the ratio of the transmitted frequency to the received frequency at the spacecraft. $\rho$ is the time interval between the reception time, referred to the electronics of the receiving ground station and expressed in Station Time\footnote{Station Time is a realization of the proper time at the Earth ground station.} (ST), and the corresponding transmission time, at the electronics of the transmitting ground station, in ST:
\begin{equation}
    \rho(t_3(\text{ST})_R) = t_3(\text{ST})_R - t_1(\text{ST})_T
\end{equation}

Neglecting second order terms, such as relativistic light-time delays, time scale transformations, electronic delays, and transmission media delays, the round-trip light-time can be written as:
\begin{equation}\label{eq:rtlt}
    \rho(t_3(\text{ST})_R) \simeq r_{23}/c + r_{12}/c = 1/c \cdot \left[ \left( \mathbf{r_3} - \mathbf{r_2} \right) \cdot \left( \mathbf{r_3} - \mathbf{r_2} \right) \right]^{1/2} + 1/c \cdot \left[ \left( \mathbf{r_2} - \mathbf{r_1} \right) \cdot \left( \mathbf{r_2} - \mathbf{r_1} \right) \right]^{1/2}
\end{equation}
where all state vectors are expressed in the Solar System barycentric space-time Frame Of Reference (FOR).

In Eq.~\ref{eq:rtlt} the position vectors $\mathbf{r_3}$ and $\mathbf{r_1}$ are computed from the planetary and lunar ephemeris and from the Earth-centered position of the ground station antenna:
\begin{gather}
    \mathbf{r_3} = \mathbf{r^C_{A3}}(t_3) = \mathbf{r^C_B}(t_3) - \mathbf{r^E_M}(t_3)/( 1 + \mu_E/\mu_M) + \mathbf{r^E_{A3}}(t_3) \\
    \mathbf{r_1} = \mathbf{r^C_{A1}}(t_1) = \mathbf{r^C_B}(t_1) - \mathbf{r^E_M}(t_1)/( 1 + \mu_E/\mu_M ) + \mathbf{r^E_{A1}}(t_1)
\end{gather}
Superscripts and subscripts designating location are explained in Table~\ref{tab:sup_sub}.

\begin{table}
\caption{Abbreviations for all bodies participating in the light-time problem.}\label{tab:sup_sub}
\begin{ruledtabular}
\begin{tabular}{cl}
Superscript/Subscript   &   Body   \\\hline
C                       &   Solar System Barycenter  \\
S                       &   Sun  \\
B                       &   Earth-Moon system barycenter  \\
M                       &   Moon  \\
E                       &   Earth  \\
A1                      &   Transmitting ground station's antenna  \\
A3                      &   Receiving ground station's antenna  \\
P                       &   Probe  \\
O                       &   Center Of Integration of the spacecraft trajectory  \\
\end{tabular}
\end{ruledtabular}
\end{table}

$\mathbf{r_2}$ is computed from the spacecraft ephemeris, obtained by integrating the equations of motion with respect to the Center Of Integration (COI), and from the planetary ephemeris:
\begin{equation}
    \mathbf{r_2} = \mathbf{r^C_P}(t_2) = \mathbf{r^C_O}(t_2) + \mathbf{r^O_P}(t_2)
\end{equation}

The reception time in Ephemeris Time\footnote{Ephemeris Time is the coordinate time of the Solar System barycentric space-time frame of reference used in the adopted celestial ephemeris.} ($T_{eph}$)~\cite{Standish:1998}, $t_3$, is computed from the reception time in ST, $t_3(\text{ST})$, through different time transformations while $t_2$ and $t_1$ are computed from $t_3$ solving for the down-link and the up-link light-time problems, respectively.

\subsection{Numerical Error Preliminary Model}\label{subsec:model_prelim}
Using the DRD formulation, the Doppler observable is computed as the difference of two round-trip light-times (Eq.~\ref{eq:doppler}). Qualitatively, this formulation has a high sensitivity to round-off errors because, as it is well known, the difference between two large and nearly equal values substantially increases the relative error, causing a loss of significance. Hence, errors in the round-trip light-time have a large influence on the Doppler observables.

At first approximation, the errors $\Delta \rho$ in the round-trip light-time at the start and at the end of the count time can be considered independent. From Eq.~\ref{eq:doppler}, an order of magnitude estimation of the corresponding effect in the Doppler observable is:
\begin{equation}\label{eq:doppler_err_from_range}
    \Delta F_{2,3} \simeq M_2 f_T \sqrt{2} \Delta \rho/T_c
\end{equation}

The main sources of numerical errors in the round-trip light-time are:
\begin{enumerate}
    \item[1)] Representation of times: the rounding errors in time epochs $t_3$, $t_2$, and $t_1$ propagate only indirectly into the round-trip light-time. An error $\Delta t_k$ in time epoch $t_k$ affects the computation of a position vector $\mathbf{r^i_j}(t_k)$, through the corresponding velocity vector $\mathbf{\dot{r}^i_j}$:
    \begin{equation}
        \Delta \mathbf{r^i_j} = \mathbf{\dot{r}^i_j} \, \Delta t_k
    \end{equation}
    At first approximation, the total effect on $\rho$ is:
    \begin{equation}\label{eq:range_err_from_time}
        \Delta \rho \simeq (\dot{r}_{12}+\dot{r}_{23}) \Delta t_k /c
    \end{equation}
    Where $\dot{r}_{ij}$ is the range-rate, i.e. the time variation of the range, between $i$ and $j$.
    
    Currently both the ODP and AMFIN express the time variable $t_k$ as a single double-precision value of the time elapsed since a reference epoch. In the ODP, time is measured in seconds past J2000 (January 1st 2000, 12:00)~\cite{Moyer:2000}, while in AMFIN, time is measured in days past January 1st 2000, 00:00~\cite{AMFIN:ALG}.
    Fig.~\ref{fig:E_t3_odpamfin} shows the maximum absolute round-off error in time variables, for both the double-precision time representations adopted in the ODP and AMFIN. When the time variable increases relative to the reference epoch, the maximum rounding error increases with a piecewise trend, because it doubles when the binary order of magnitude $p(t)$ increases by one.
    \begin{figure}
        \centering
        \includegraphics[width=\textwidth]{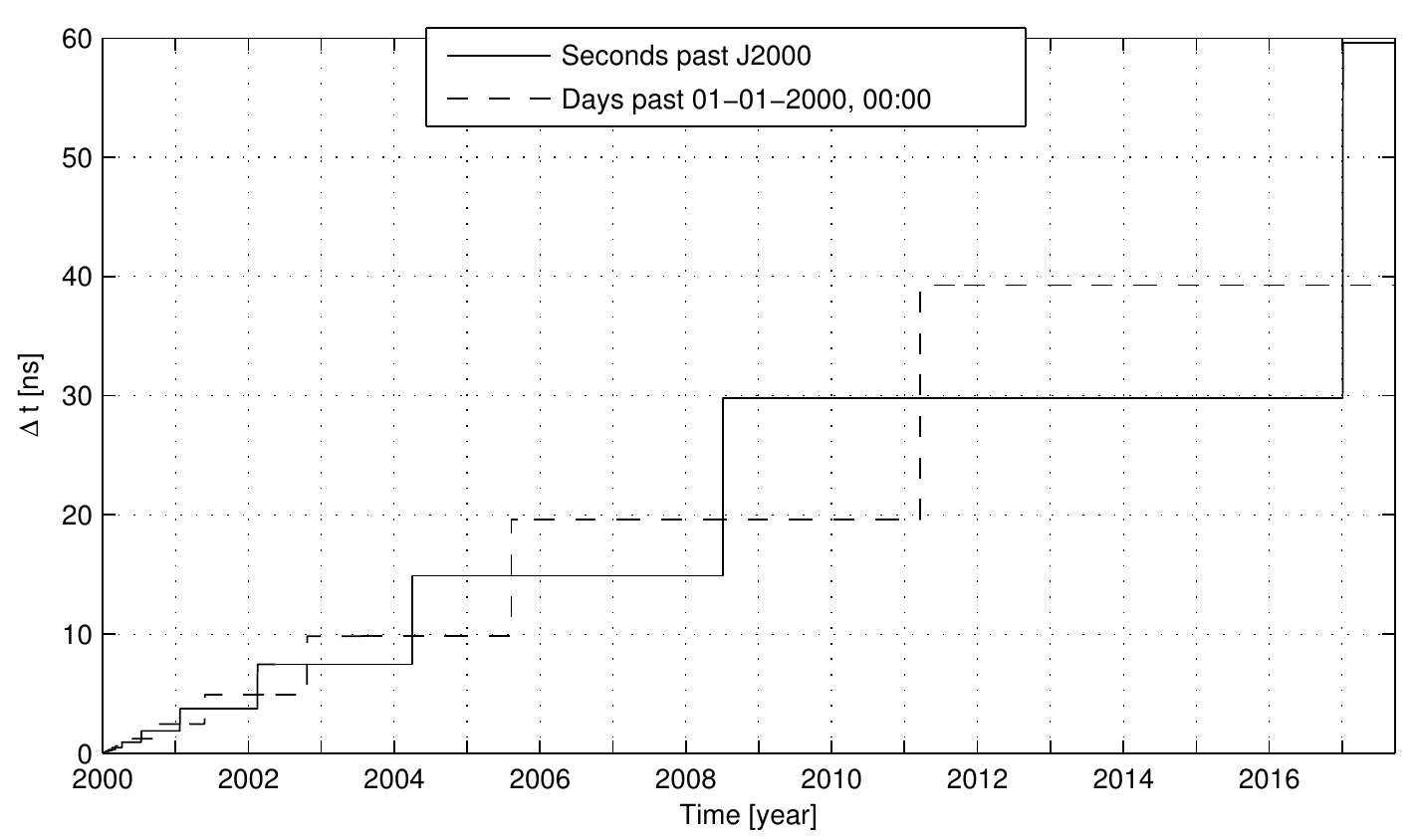}
        \caption{\label{fig:E_t3_odpamfin} Maximum absolute round-off errors in time variable for the two double-precision time representation adopted in the ODP (seconds past J2000) and AMFIN (days past January 1st 2000, 00:00).}
    \end{figure}
    The difference of 12 hours in the reference epochs of the two programs is negligible. Hence, as discussed in Sec.~\ref{sec:num_err_theory}, because of the different measurement units, the time round-off errors in these two OD codes are not exactly the same, but they have the same average magnitude, on a large enough time interval.
    For the ODP-like time representation, between July 2008 and January 2017, the maximum round-off error is about $30\, \text{ns}$. From Eq.~\ref{eq:range_err_from_time}, considering a range-rate of $30\,\text{km/s}$, the effect of the time error on the two-way range has an order of magnitude of $2 \, \text{mm}$. From Eq.~\ref{eq:doppler_err_from_range}, the approximate effect on the two-way range-rate is about $4 \times 10^{-2} \,\text{mm/s}$, for a count time of $60 \, \text{s}$.
    \item[2)] Representation of distances: the rounding errors in each component of the position vectors propagate into the precision round-trip light-time directly, because the position vectors are used in the computation of the Newtonian distances $r_{12}$ and $r_{23}$, but also indirectly, because the position vectors are used also in the computation of $t_2$ and $t_1$. However, the indirect effects are much smaller than the direct ones and can be neglected.
    Both the ODP and AMFIN express distances in km, so at $1 \,\text{AU}$ the maximum round-off error in the two-way range is about $0.03 \, \text{mm}$ and at $10 \,\text{AU}$ it is about $0.24 \, \text{mm}$. From Eq.~\ref{eq:doppler_err_from_range}, the approximate effect on the two-way range-rate is about $7 \times 10^{-4} \,\text{mm/s}$ at $1 \,\text{AU}$ and $6 \times 10^{-3}\,\text{mm/s}$ at $10 \,\text{AU}$, for a count time of $60 \, \text{s}$.
    \item[3)] Additional rounding errors: the rounding errors in the result of each operation propagate into the round-trip light-time. In the computation of the round-trip light-time $\rho$, the following computational steps were considered:
    \begin{gather}
        \mathbf{r^E_B}(t_i) = \mathbf{r^E_M}(t_i)/(1 + \mu_E / \mu_M),                    \qquad i=1,3    \label{eq:reb}  \\
        \mathbf{r^C_E}(t_i) = \mathbf{r^C_B}(t_i) - \mathbf{r^E_B}(t_i),                        \qquad i=1,3    \label{eq:rce}  \\
        \mathbf{r_i} = \mathbf{r^C_{Ai}}(t_i) = \mathbf{r^C_E}(t_i) + \mathbf{r^E_{Ai}}(t_i),   \qquad i=1,3    \label{eq:ri}   \\
        \mathbf{r_2} = \mathbf{r^C_P}(t_2) = \mathbf{r^C_O}(t_2) + \mathbf{r^O_P}(t_2)                          \label{eq:r2}   \\
        \mathbf{r_{ij}} = \mathbf{r_j} - \mathbf{r_i} = \left[x_{ij} \quad y_{ij} \quad z_{ij} \right]^\text{T},    \qquad (i,j)=(1,2),(2,3)     \\
        xx_{ij} = x_{ij}^2, \quad yy_{ij} = y_{ij}^2, \quad zz_{ij} = z_{ij}^2,                                     \qquad (i,j)=(1,2),(2,3)     \\
        xy_{ij} = xx_{ij} + yy_{ij},    \qquad (i,j)=(1,2),(2,3)     \\
        rr_{ij} = xy_{ij} + zz_{ij},    \qquad (i,j)=(1,2),(2,3)     \\
        r_{ij} = \sqrt{rr_{ij}}, \qquad (i,j)=(1,2),(2,3)     \\
        t_{ij} = r_{ij}/c, \qquad (i,j)=(1,2),(2,3)     \\
        \rho = t_{12} + t_{23}
    \end{gather}
    
    Moreover, the position vectors $\mathbf{r^i_j}$ used in Eqs.~\ref{eq:reb}, \ref{eq:rce}, \ref{eq:ri}, and \ref{eq:r2} are computed using the astronomical ephemeris (given as an input to the program), the integration of the equations of motion and the Earth-fixed location of the ground station antenna. For simplicity, the numerical errors introduced in these computational steps are neglected. This assumption was verified \emph{a posteriori} through the validation of the numerical errors model.
\end{enumerate}

As discussed in Sec.~\ref{subsec:num_err_theory_oper}, a numerical error $\Delta \overline{x}$ in a generic variable $x$ propagates into the output variable $z$ through the partial derivative $\partial z / \partial x$.
Hence, the numerical noise in the round-trip light-time $\rho$ at reception time $t_3$ can be computed through the partial derivative of $\rho$ with respect to each considered error source. Neglecting second order terms, the resulting expression for the total numerical error on $\rho$ is:
\begin{multline}\label{eq:delta_rho_tot}
    \Delta \rho(t_3) = (\dot{r}_{12}+\dot{r}_{23}) \Delta \overline{t_3}/c + (\dot{r}_{12}-\dot{p}_{23}) \Delta \overline{t_2}/c - \dot{p}_{12} \Delta \overline{t_1}/c \\
    + (\mathbf{\hat{r}_{12}} - \mathbf{\hat{r}_{23}})/c \cdot \left[ \Delta\overline{\mathbf{r^O_P}}(t_2) + \Delta\overline{\mathbf{r^C_O}}(t_2) + \Delta\overline{\mathbf{r^C_P}}(t_2) \right] \\
    + \frac{\mathbf{\hat{r}_{23}}}{c} \cdot \left[ \Delta\overline{\mathbf{r^E_{A3}}}(t_3) + \Delta\overline{\mathbf{r^C_B}}(t_3)  - \frac{\Delta\overline{\mathbf{r^E_M}}(t_3)}{1 + \mu_E / \mu_M} + \Delta\overline{\mathbf{r^C_{A3}}}(t_3) + \Delta\overline{\mathbf{r^C_E}}(t_3)  - \Delta\overline{\mathbf{r^E_B}}(t_3) + \Delta\overline{\mathbf{r_{23}}} \right] \\
    - \frac{\mathbf{\hat{r}_{12}}}{c} \cdot \left[ \Delta\overline{\mathbf{r^E_{A3}}}(t_1) + \Delta\overline{\mathbf{r^C_B}}(t_1) - \frac{\Delta\overline{\mathbf{r^E_M}}(t_1)}{1 + \mu_E / \mu_M} + \Delta\overline{\mathbf{r^C_{A1}}}(t_1) + \Delta\overline{\mathbf{r^C_E}}(t_1)  - \Delta\overline{\mathbf{r^E_B}}(t_1) - \Delta\overline{\mathbf{r_{12}}} \right] \\
    + \frac{\Delta\overline{xx_{12}} + \Delta\overline{yy_{12}} + \Delta\overline{zz_{12}} + \Delta\overline{xy_{12}} + \Delta\overline{rr_{12}}}{2\,c\,r_{12}} + \frac{\Delta\overline{xx_{23}} + \Delta\overline{yy_{23}} + \Delta\overline{zz_{23}} + \Delta\overline{xy_{23}} + \Delta\overline{rr_{23}}}{2\,c \,r_{23}} \\
    + \left( \Delta\overline{r_{12}} + \Delta\overline{r_{23}} \right) /c + \Delta\overline{t_{12}} + \Delta\overline{t_{23}} + \Delta\overline{\rho}
\end{multline}
where $\dot{p}_{ij}$ is defined as:
\begin{equation}
    \dot{p}_{ij} \triangleq \mathbf{\dot{r}_{i}} \cdot \mathbf{\hat{r}_{ij}}     \qquad (i,j)=(1,2),(2,3)     \\
\end{equation}

The numerical error in two- or three-way Doppler observables with time-tag $TT$ can be expressed as:
\begin{equation}\label{eq:delta_dop}
    \Delta F_{2,3}(TT) = M_2 f_T \left[ \Delta \rho(TT+T_c/2) - \Delta \rho(TT-T_c/2) \right] /T_c
\end{equation}

Adopting the statistical model for each numerical error in Eq.~\ref{eq:delta_rho_tot}, it is possible to evaluate the statistical properties of the total numerical error in the round-trip light-time $\rho$ and Doppler observable $F_{2,3}$. It follows that $\Delta\rho(t_3)$ is a stochastic process with the following characteristics:
\begin{enumerate}
    \item[1)] White.
    \item[2)] Non stationary, because the coefficients multiplying every single numerical error are a function of time. However their time variation is typically very slow and, for a typical duration of a tracking pass, the process can be considered stationary.
    \item[3)] Gaussian-like distribution/PDF: from the central limit theorem the sum of $N$ independent and identically distributed random variables converges to a normally distributed random variable as $N$ approaches infinity. Because the different numerical errors are a finite number and not identically distributed, the expected PDF is only qualitatively bell-shaped.
    \item[4)] Zero mean.
    \item[5)] Variance: the total variance is the sum of each term's variance:
    \begin{multline}\label{eq:sigma_delta_rho_tot}
        \sigma^2_{\Delta \rho}(t_3) = \left( \frac{\dot{r}_{12}+\dot{r}_{23}}{c} \right)^2 \sigma^2_{\Delta \overline{t_3}} + \left( \frac{\dot{r}_{12}-\dot{p}_{23}}{c} \right)^2 \sigma^2_{\Delta \overline{t_2}} + \left( \frac{\dot{p}_{12}}{c} \right)^2 \sigma^2_{\Delta \overline{t_1}} \\
        + \left( \frac{x_{12}/r_{12} - x_{23}/r_{23}}{c} \right)^2 \left[ \sigma^2_{\Delta\overline{x^O_P}}(t_2) + \sigma^2_{\Delta\overline{x^C_O}}(t_2) + \sigma^2_{\Delta\overline{x^C_P}}(t_2) \right] \\
        + \left( \frac{y_{12}/r_{12} - y_{23}/r_{23}}{c} \right)^2 \left[ \sigma^2_{\Delta\overline{y^O_P}}(t_2) + \sigma^2_{\Delta\overline{y^C_O}}(t_2) + \sigma^2_{\Delta\overline{y^C_P}}(t_2) \right] \\
        + \left( \frac{z_{12}/r_{12} - z_{23}/r_{23}}{c} \right)^2 \left[ \sigma^2_{\Delta\overline{z^O_P}}(t_2) + \sigma^2_{\Delta\overline{z^C_O}}(t_2) + \sigma^2_{\Delta\overline{z^C_P}}(t_2) \right] \\
        + \left( \frac{x_{23}}{c\,r_{23}} \right)^2 \left[ \sigma^2_{\Delta\overline{x^E_{A3}}}(t_3) + \sigma^2_{\Delta\overline{x^C_B}}(t_3)  + \frac{\sigma^2_{\Delta\overline{x^E_M}}(t_3)}{(1 + \mu_E / \mu_M)^2} + \sigma^2_{\Delta\overline{x^C_{A3}}}(t_3) + \sigma^2_{\Delta\overline{x^C_E}}(t_3)  + \sigma^2_{\Delta\overline{x^E_B}}(t_3) + \sigma^2_{\Delta\overline{x_{23}}} \right] \\
        + \left( \frac{y_{23}}{c\,r_{23}} \right)^2 \left[ \sigma^2_{\Delta\overline{y^E_{A3}}}(t_3) + \sigma^2_{\Delta\overline{y^C_B}}(t_3)  + \frac{\sigma^2_{\Delta\overline{y^E_M}}(t_3)}{(1 + \mu_E / \mu_M)^2} + \sigma^2_{\Delta\overline{y^C_{A3}}}(t_3) + \sigma^2_{\Delta\overline{y^C_E}}(t_3)  + \sigma^2_{\Delta\overline{y^E_B}}(t_3) + \sigma^2_{\Delta\overline{y_{23}}} \right] \\
        + \left( \frac{z_{23}}{c\,r_{23}} \right)^2 \left[ \sigma^2_{\Delta\overline{z^E_{A3}}}(t_3) + \sigma^2_{\Delta\overline{z^C_B}}(t_3)  + \frac{\sigma^2_{\Delta\overline{z^E_M}}(t_3)}{(1 + \mu_E / \mu_M)^2} + \sigma^2_{\Delta\overline{z^C_{A3}}}(t_3) + \sigma^2_{\Delta\overline{z^C_E}}(t_3)  + \sigma^2_{\Delta\overline{z^E_B}}(t_3) + \sigma^2_{\Delta\overline{z_{23}}} \right] \\
        + \left( \frac{x_{12}}{c\,r_{12}} \right)^2 \left[ \sigma^2_{\Delta\overline{x^E_{A3}}}(t_1) + \sigma^2_{\Delta\overline{x^C_B}}(t_1) + \frac{\sigma^2_{\Delta\overline{x^E_M}}(t_1)}{(1 + \mu_E / \mu_M)^2} + \sigma^2_{\Delta\overline{x^C_{A1}}}(t_1) + \sigma^2_{\Delta\overline{x^C_E}}(t_1)  + \sigma^2_{\Delta\overline{x^E_B}}(t_1) + \sigma^2_{\Delta\overline{x_{12}}} \right] \\
        + \left( \frac{y_{12}}{c\,r_{12}} \right)^2 \left[ \sigma^2_{\Delta\overline{y^E_{A3}}}(t_1) + \sigma^2_{\Delta\overline{y^C_B}}(t_1) + \frac{\sigma^2_{\Delta\overline{y^E_M}}(t_1)}{(1 + \mu_E / \mu_M)^2} + \sigma^2_{\Delta\overline{y^C_{A1}}}(t_1) + \sigma^2_{\Delta\overline{y^C_E}}(t_1)  + \sigma^2_{\Delta\overline{y^E_B}}(t_1) + \sigma^2_{\Delta\overline{y_{12}}} \right] \\
        + \left( \frac{z_{12}}{c\,r_{12}} \right)^2 \left[ \sigma^2_{\Delta\overline{z^E_{A3}}}(t_1) + \sigma^2_{\Delta\overline{z^C_B}}(t_1) + \frac{\sigma^2_{\Delta\overline{z^E_M}}(t_1)}{(1 + \mu_E / \mu_M)^2} + \sigma^2_{\Delta\overline{z^C_{A1}}}(t_1) + \sigma^2_{\Delta\overline{z^C_E}}(t_1)  + \sigma^2_{\Delta\overline{z^E_B}}(t_1) + \sigma^2_{\Delta\overline{z_{12}}} \right] \\
        + \frac{\sigma^2_{\Delta\overline{xx_{12}}} + \sigma^2_{\Delta\overline{yy_{12}}} + \sigma^2_{\Delta\overline{zz_{12}}} + \sigma^2_{\Delta\overline{xy_{12}}} + \sigma^2_{\Delta\overline{rr_{12}}}}{(2\,c\,r_{12})^2} + \frac{\sigma^2_{\Delta\overline{xx_{23}}} + \sigma^2_{\Delta\overline{yy_{23}}} + \sigma^2_{\Delta\overline{zz_{23}}} + \sigma^2_{\Delta\overline{xy_{23}}} + \sigma^2_{\Delta\overline{rr_{23}}}}{(2\,c\,r_{23})^2} \\
        + \left( \sigma^2_{\Delta\overline{r_{12}}} + \sigma^2_{\Delta\overline{r_{23}}} \right)/c^2 + \sigma^2_{\Delta\overline{t_{12}}} + \sigma^2_{\Delta\overline{t_{23}}} + \sigma^2_{\Delta\overline{\rho}}
    \end{multline}
    Where the variance of each numerical error can be computed using the formulation presented in Sec.~\ref{subsec:stat_mod}.
\end{enumerate}

The numerical error $\Delta F_{2,3}$ has similar characteristics to $\Delta\rho$ but it is not white, because two consecutive Doppler observables are a function of the round-trip light-time at their mid-time.
In fact, the count times are consecutive, so the end of a count time is the start of the next count time:
\begin{equation}\label{eq:count_times}
    TT_{k}+T_c/2 = TT_{k+1}-T_c/2
\end{equation}
Given Eq.~\ref{eq:count_times}, two consecutive Doppler observables can be written as:
\begin{align}
    F_{2,3}(TT_{k})   &= M_2 f_T \left[ \rho(TT_{k}+T_c/2) -\rho(TT_{k}-T_c/2) \right] /T_c \notag \\
                      &= M_2 f_T \left[ \rho(TT_{k+1}-T_c/2) -\rho(TT_{k}-T_c/2) \right] /T_c \\
    F_{2,3}(TT_{k+1}) &= M_2 f_T \left[ \rho(TT_{k+1}+T_c/2) -\rho(TT_{k+1}-T_c/2) \right] /T_c \notag \\
                      &= M_2 f_T \left[ \rho(TT_{k+2}-T_c/2) -\rho(TT_{k+1}-T_c/2) \right] /T_c
\end{align}
Hence, the rounding error in $\rho(TT_{k+1}-T_c/2)$ affects both $F_{2,3}(TT_{k})$ and $F_{2,3}(TT_{k+1})$. The autocorrelation function of $\Delta F_{2,3}$ is:
\begin{multline}
    R_{\Delta F}(\tau) \triangleq \frac{\E \left[ \left( \Delta F(t) \right) \left( \Delta F(t+\tau) \right) \right]}{\sigma_{\Delta F(t)} \sigma_{\Delta F(t+\tau)}} \\
    \simeq \frac{2 R_{\Delta \rho}(\tau) - R_{\Delta \rho}(\tau - T_c) - R_{\Delta \rho}(\tau + T_c)}{2 \left[ 1 - R_{\Delta \rho}(T_c) \right]} =
    \begin{cases}
        1,      &   \text{if $\tau = 0$,} \\
        -1/2,   &   \text{if $|\tau| = T_c$,} \\
        0,      &   \text{if $|\tau| > T_c$}.
    \end{cases}
\end{multline}

The variance of $\Delta F_{2,3}$ is proportional to the variance of $\Delta \rho$:
\begin{multline}\label{sigma_delta_dop2_tot}
    \sigma^2_{\Delta F_{2,3}}(TT) = \left( M_2 f_T/T_c \right)^2 \left[ \sigma^2_{\Delta \rho}(TT+T_c/2) + \sigma^2_{\Delta \rho}(TT-T_c/2) \right] \simeq 2 \left( M_2 f_T \sigma_{\Delta \rho}(TT) / T_c \right)^2
\end{multline}

Thus, the standard deviation of $\Delta F_{2,3}$, representing the numerical noise level in Doppler observables, depends on:
\begin{enumerate}
    \item[1)] Floating point word's length: the number of bits used for the fractional part of the mantissa defines the amplitude of all round-off errors. At present, almost all hardware and compilers implement the IEEE-754 standard.
    \item[2)] Time: the time at which the OD problem is settled defines the order of magnitude of all time epochs, thus defining the quantization step and the amplitude of the round-off errors in each time variable.
    \item[3)] Geometry: the state vectors of all participants to the OD problem define two critical factors: the order of magnitude of the three components of each position vector, from which the range and the additional round-off errors derive, and the velocities of the participants that define the influence of time round-off errors in the observables.
    \item[4)] Count time: $\Delta\rho$ does not depend upon $T_c$, hence $\Delta F_{2,3}$ (and its standard deviation) varies with $T_c^{-1}$. As a reference a typical white-phase reference noise varies with $T_c^{-1/2}$.
\end{enumerate}

\subsection{Numerical Error Complete Model}\label{subsec:model_complete}
The most delicate assumption on which the numerical noise model is based is the statistical characterization of the round-off errors using Widrow's model.
In order to verify this assumption and to study in detail the numerical errors, a Radiometric Observables Generator (ROG) computer code was developed. The program computes the two-way Doppler observables on the basis of Moyer's DRD formulation, with the following simplifications:
\begin{enumerate}
    \item[1)] The planetary ephemeris and the space-fixed state of the ground station were computed using the SPICE kernels and toolkit~\cite{Acton:1996}.
    \item[2)] The spacecraft trajectory was computed using the ODP, considering only the gravitational accelerations.
    \item[3)] In the solution of the light-time problems and in the computation of the precision light-times only the Newtonian terms were considered.
\end{enumerate}

To study the numerical errors in a realistic scenario, the spacecraft trajectory was integrated from the real state vector of ESA's Rosetta interplanetary spacecraft from January 1, 2008\footnote{The spacecraft state was retrieved retrieved from the latest Rosetta's SPICE kernels, available at \url{ftp://psa.esac.esa.int/pub/mirror/INTERNATIONAL-ROSETTA-MISSION/SPICE/ROS-E-M-A-C-SPICE-6-V1.0/DATA/SPK/} [last accessed October 21, 2011].}, to December 31, 2010. To simplify the simulation only gravitational accelerations were considered, neglecting all non-gravitational effects. Hence the ODP did not reproduce the real Rosetta trajectory, and in the following this scenario will be referred to as the ``Rosetta-like scenario''.

All steps in the computations were performed using both the quadruple and double-precision floating point representation. Because the quadruple-precision machine epsilon is about $10^{18}$ times smaller than the double-precision one, the numerical error in the quadruple-precision values can be neglected and the computed quantities can be considered as infinitely precise reference values. Hence, a very good estimation of the ``real numerical errors'' in each double-precision quantity can be obtained as the difference between the double-precision values and the quadruple-precision ones.

\begin{figure}
\centering
\includegraphics[width=\textwidth]{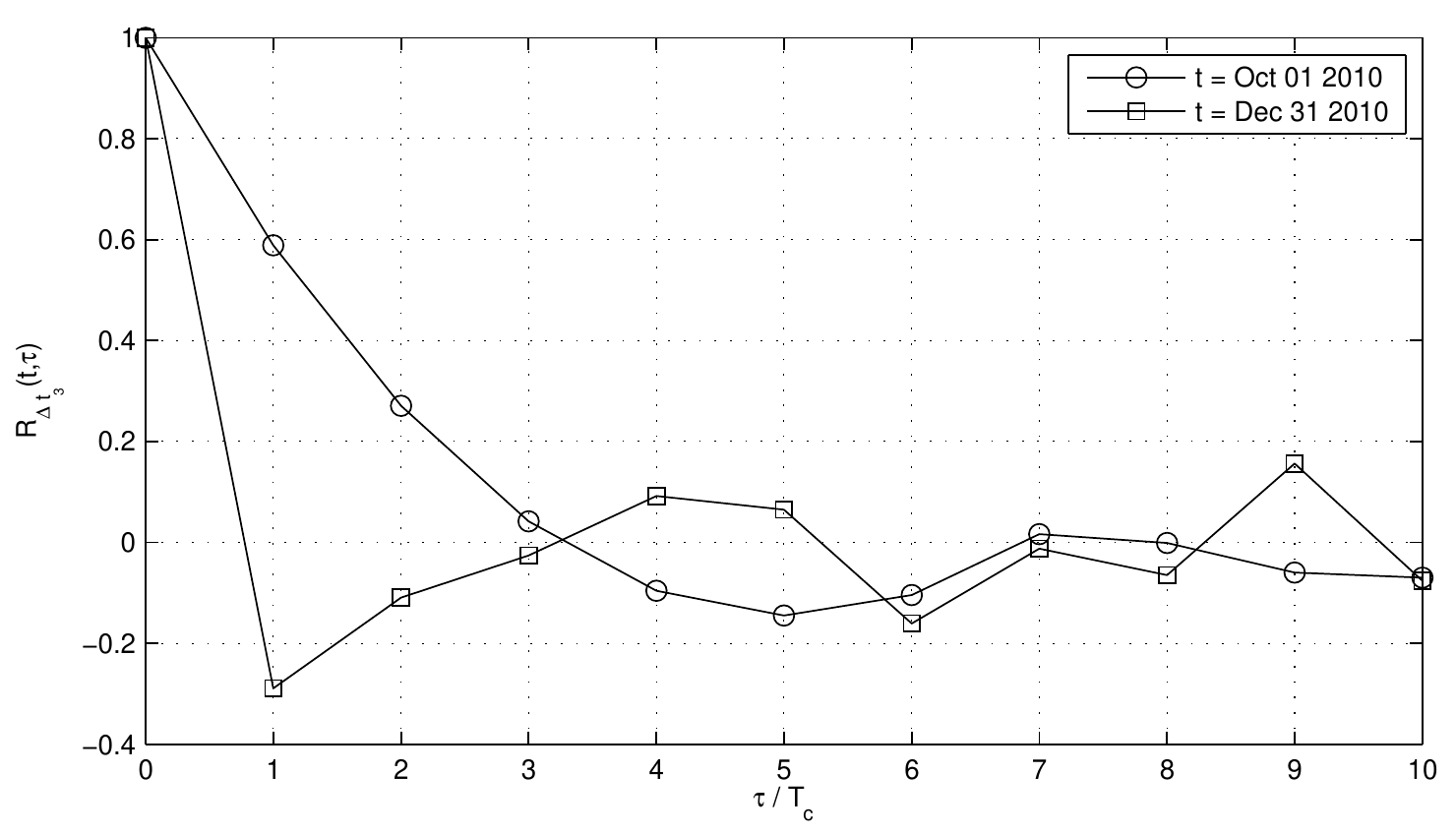}
\caption{\label{fig:auto_e_t3_pass} Autocorrelation of $\Delta \overline{t_3}$, computed on different tracking passes.}
\end{figure}

The ROG program was used to study all elementary round-off errors appearing in Eq.~\ref{eq:delta_rho_tot}.
As a result of the analysis, Widrow's model was proven to be a good description of all round-off errors but $\Delta \overline{t_3}$, which cannot be considered a white random process, because its autocorrelation $R_{\Delta\overline{t_3}}(t,T_c)$ is not a Dirac delta function.
An example of the autocorrelation of $\Delta \overline{t_3}$, computed at two different passes, is shown in Fig.~\ref{fig:auto_e_t3_pass}.
This error can still be described using an immediate extension of Widrow's model, obtained neglecting the whiteness property: $\Delta \overline{t_3}$ is modeled as a non-white, wide-sense stationary (on a single pass), uniformly distributed with zero mean, random process.

It can be shown that the numerical error in $t_3$ is the rounding error of the quantity $(T_{eph}-\text{TAI})_{t_3}$ considering the quantization step of $t_3$:
\begin{equation}\label{eq:delta_t3_intero}
    \Delta \overline{t_3} = \round \left[ (T_{eph}-\text{TAI})_{t_3}/q(t_3) \right] \cdot q(t_3) - (T_{eph}-\text{TAI})_{t_3}
\end{equation}
where $(T_{eph}-\text{TAI})_t$ is the difference between the timescales $T_{eph}$ and International Atomic Time (TAI)~\cite{Terrien:1971,Terrien:1975} at ephemeris time $t$. It is given by a number of relativistic terms and it is mainly a function of the Earth's orbital motion around the Sun. Hence, this time difference is approximately a sinusoid with mean value $32.184 \, \text{s}$, peak-to-peak amplitude of about $3.3 \, \text{ms}$ and period of nearly $1$ year. Moreover, $\Delta \overline{t_3}$ is also a function of the quantization step of $t_3$, $q(t_3)$, that is a piecewise constant function of time.

\begin{figure}
\centering
\includegraphics[width=\textwidth]{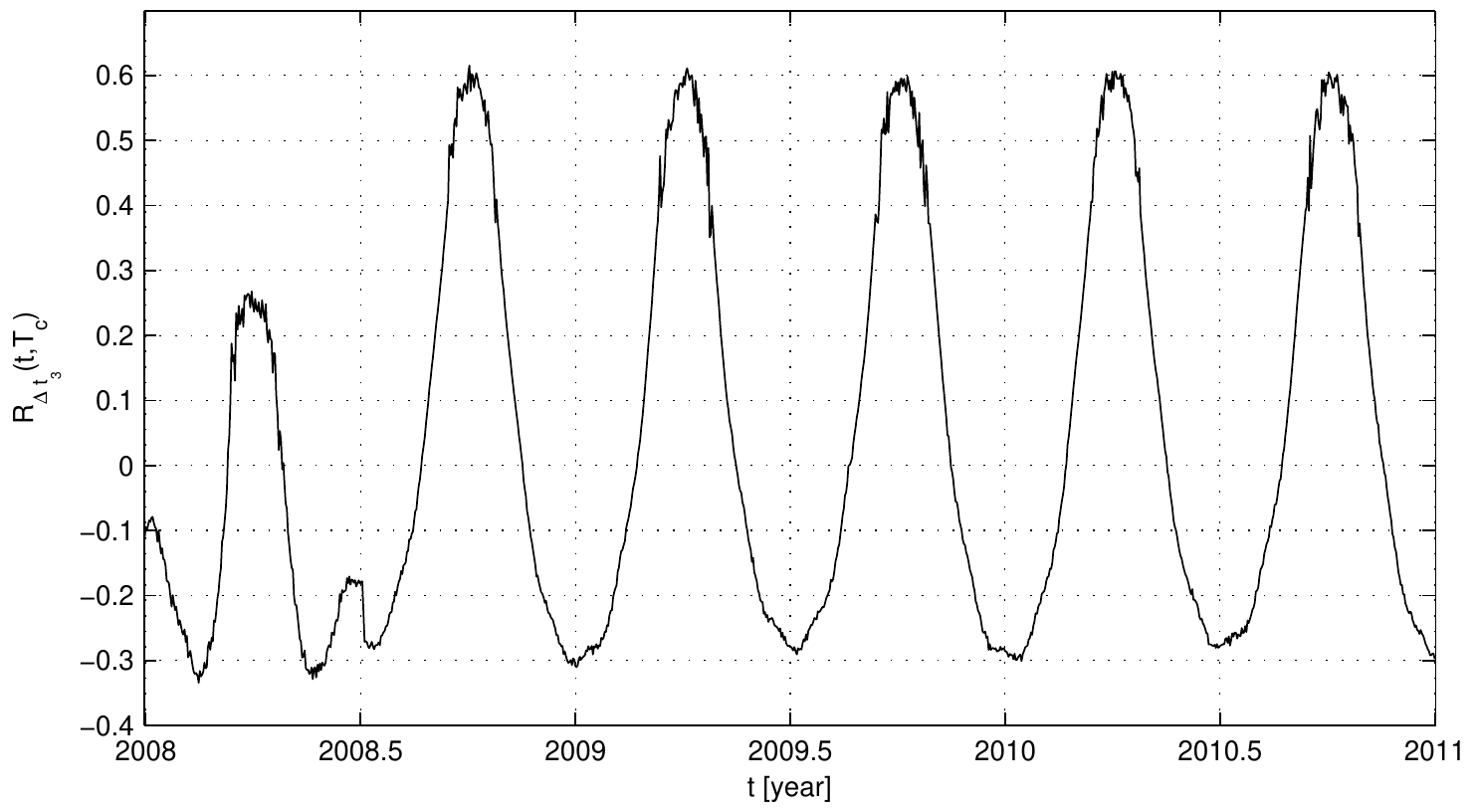}
\caption{\label{fig:auto_e_t3_time} Time variation of the autocorrelation of $\Delta \overline{t_3}$ after $T_c = 60 \,\text{s}$.}
\end{figure}

The autocorrelation $R_{\Delta\overline{t_3}}(t,\tau)$ can be computed using the real error $\Delta \overline{t_3}$ obtained from Eq.~\ref{eq:delta_t3_intero}. Fig.~\ref{fig:auto_e_t3_time}, representing $R_{\Delta\overline{t_3}}(t,T_c)$ using $T_c = 60 \, \text{s}$, shows a periodicity of $1/2$ year that derives from $(T_{eph}-\text{TAI})_t$ and, at about the middle of 2008, a discontinuity caused by the sudden doubling of $q(t_3)$.

Dropping the whiteness property of $\Delta \overline{t_3}$, the expected statistical characteristics of $\Delta\rho$ and $\Delta F_{2,3}$ change.
$\Delta \rho$ has the same properties described in Sec.~\ref{subsec:model_prelim}, except that it is a non-white random process, and its autocorrelation is a function of the autocorrelation of $\Delta \overline{t_3}$:
\begin{equation}
    R_{\Delta\rho}(t,\tau) = \left[ \left( \dot{r}_{12}+\dot{r}_{23} \right)/c \right]^2 \cdot \sigma^2_{\Delta \overline{t_3}}/\sigma^2_{\Delta\rho} \cdot R_{\Delta \overline{t_3}}(t,\tau)
\end{equation}

$\Delta F_{2,3}$ is a random process with the following characteristics:
\begin{enumerate}
    \item[1)] Non-white, with a non-stationary autocorrelation $R_{\Delta F_{2,3}}(t,\tau)$.
    \item[2)] Distribution: because of the autocorrelation of $\Delta \overline{t_3}$, the distribution may be very different from a bell-shaped curve. The expected probability density function of $\Delta F_{2,3}$ was not computed in this study, due its complexity and its very low practical utility.
    \item[3)] Zero Mean.
    \item[4)] ASDEV: the expected Allan standard deviation $\sigma_y(\tau)$ is proportional to $\tau^{-1}$.
    \item[5)] Standard Deviation: Eq.~\ref{sigma_delta_dop2_tot} must be changed to account for the autocorrelation of $\Delta \overline{t_3}$ after $T_c$ seconds:
    \begin{multline}
        \sigma_{\Delta F}^2(TT) \simeq \left( M_2 f_T/T_c \right)^2 \Biggl[ 2 \, \sigma_{\Delta \rho}^2(TT) \\
        - 2 \Bigl[ \dot{r}_{12}(TT)+\dot{r}_{23}(TT) \Bigr]^2 \cdot \sigma^2_{\Delta \overline{t_3}}(TT) \cdot R_{\Delta \overline{t_3}}(TT,T_c)/c^2 \Biggr]
    \end{multline}
\end{enumerate}

\section{Model Validation}\label{sec:model_validation}
The complete numerical error model for Doppler observables described in Sec.~\ref{subsec:model_complete} was validated by analyzing the numerical noise in the Doppler observables computed by NASA/JPL's ODP. The ODP is a state-of-the-art interplanetary OD program, developed since the 1960s at NASA's JPL, and it has been used for navigation in almost all NASA missions controlled by JPL~\cite{Tortora:2002, Tortora:2004, Bertotti:2008}.

For the purpose of this study, the ODP was used to replicate the ``Rosetta-like scenario'', described in Sec.~\ref{subsec:model_complete}, integrating the trajectory in the same cruise phase and computing the corresponding two-way Doppler observables with respect to an Earth ground station.

The numerical noise was extracted from the computed observables using the so-called ``six-parameter fit'', a well known and simple method usually employed to estimate the noise content of Doppler observed observables~\cite{Curkendall:1969, Bertotti:1995}.
The fit is based upon a simplified formulation of the range-rate between the spacecraft and the Earth ground station:
\begin{equation}\label{eq:range_rate_simple}
    \dot{r}^{Ai}_{P}(t) = \dot{r}^{E}_{P}(t) + \omega_e r_s \cos{\delta(t)}\sin{\left( \omega_e t + \alpha_0 + \lambda_s - \alpha(t) \right)},  \qquad i=1,3
\end{equation}

To take into account the Earth-spacecraft relative motion during the tracking pass, $\dot{r}^{E}_{P}(t)$, $\delta(t)$, and $\alpha(t)$ are expanded to a first-order Taylor time series centered at the start of the pass. Then, following~\cite{Curkendall:1969}, from Eq.~\ref{eq:range_rate_simple} the range-rate can be expressed as a linear combination of the following six functions:
\begin{equation}\label{eq:spf}
    1, \quad t_p, \quad \sin{\omega_e t_p}, \quad \cos{\omega_e t_p}, \quad t_p \sin{\omega_e t_p}, \quad t_p \cos{\omega_e t_p}
\end{equation}

From Eq.~\ref{eq:doppler}, the two-way and three-way Doppler observable is proportional to the mean range-rate during the count-time. Hence, during a single tracking pass, the six functions in Eq.~\ref{eq:spf} can be used to fit $F_{2,3}$, in a least-squares sense. The six estimated parameters, which define the fitting function, are related to the spacecraft geocentric state.

Due to the first-order approximation, the six-parameter fit can be successfully applied only in the cruise phase, when the spacecraft does not experience significant accelerations. Hence, during the numerical noise extraction, tracking passes corresponding to planetary fly-bys were neglected.

In the absence of artificial additive random quantities, the numerical noise is the only non-negligible noise in the computed observables.
Hence, the analysis is based upon the assumption that the six-parameter fit is capable of approximating (in a least-squares sense) the theoretical, infinitely precise, computed Doppler observables, leaving only the numerical errors as the fit's residuals.

\begin{figure}
\centering
\includegraphics[width=\textwidth]{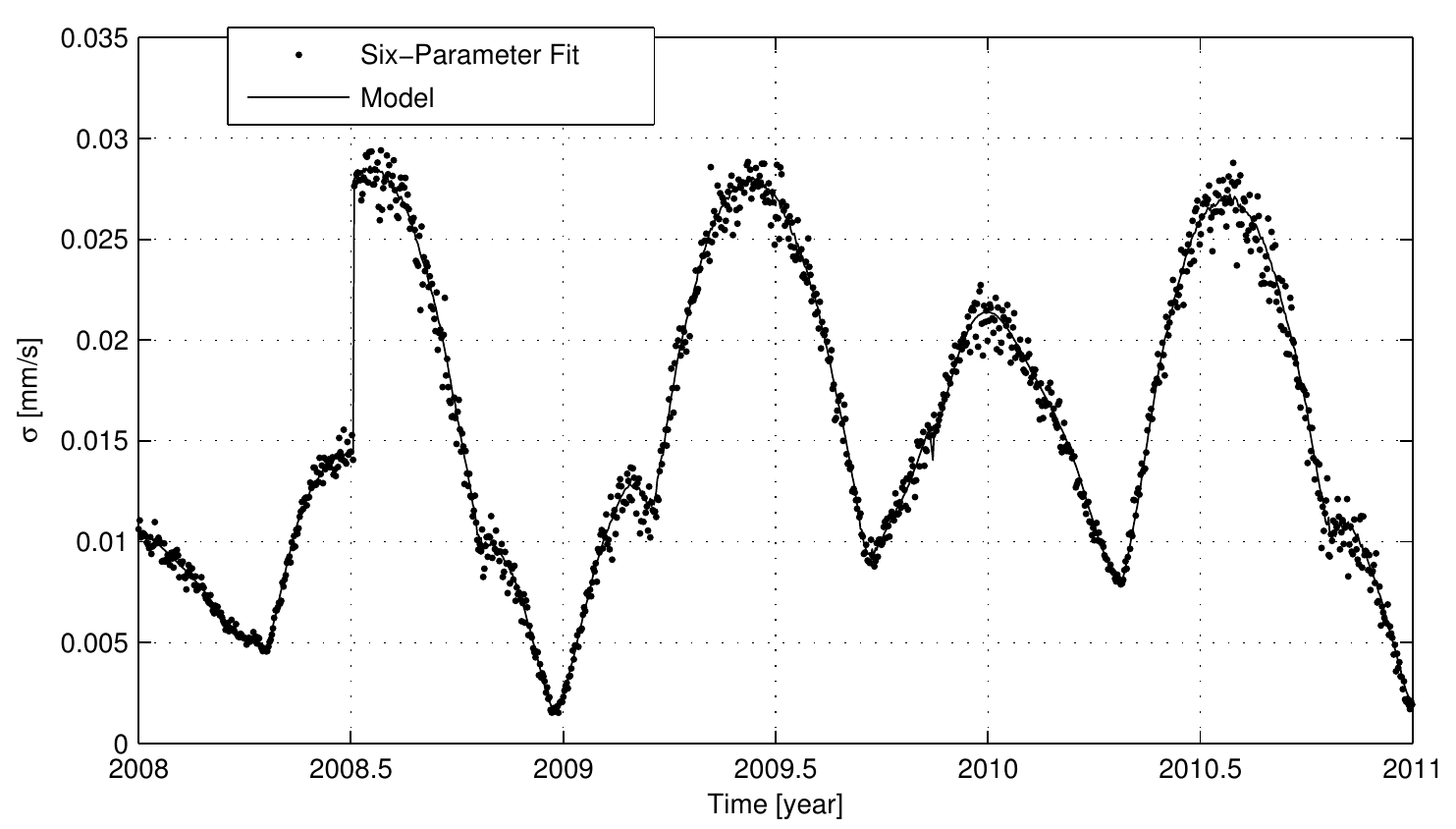}
\caption{\label{fig:sigma_spfodp_mod} Six-parameter fit applied to the Doppler observables computed by the ODP: comparison between the standard deviation of the fit's residuals, $\sigma_{ODP}$, computed on each tracking pass, and the expected standard deviation of the numerical errors, $\sigma_{MOD}$, obtained from the final version of the model.}
\end{figure}

\begin{figure}
\centering
\includegraphics[width=\textwidth]{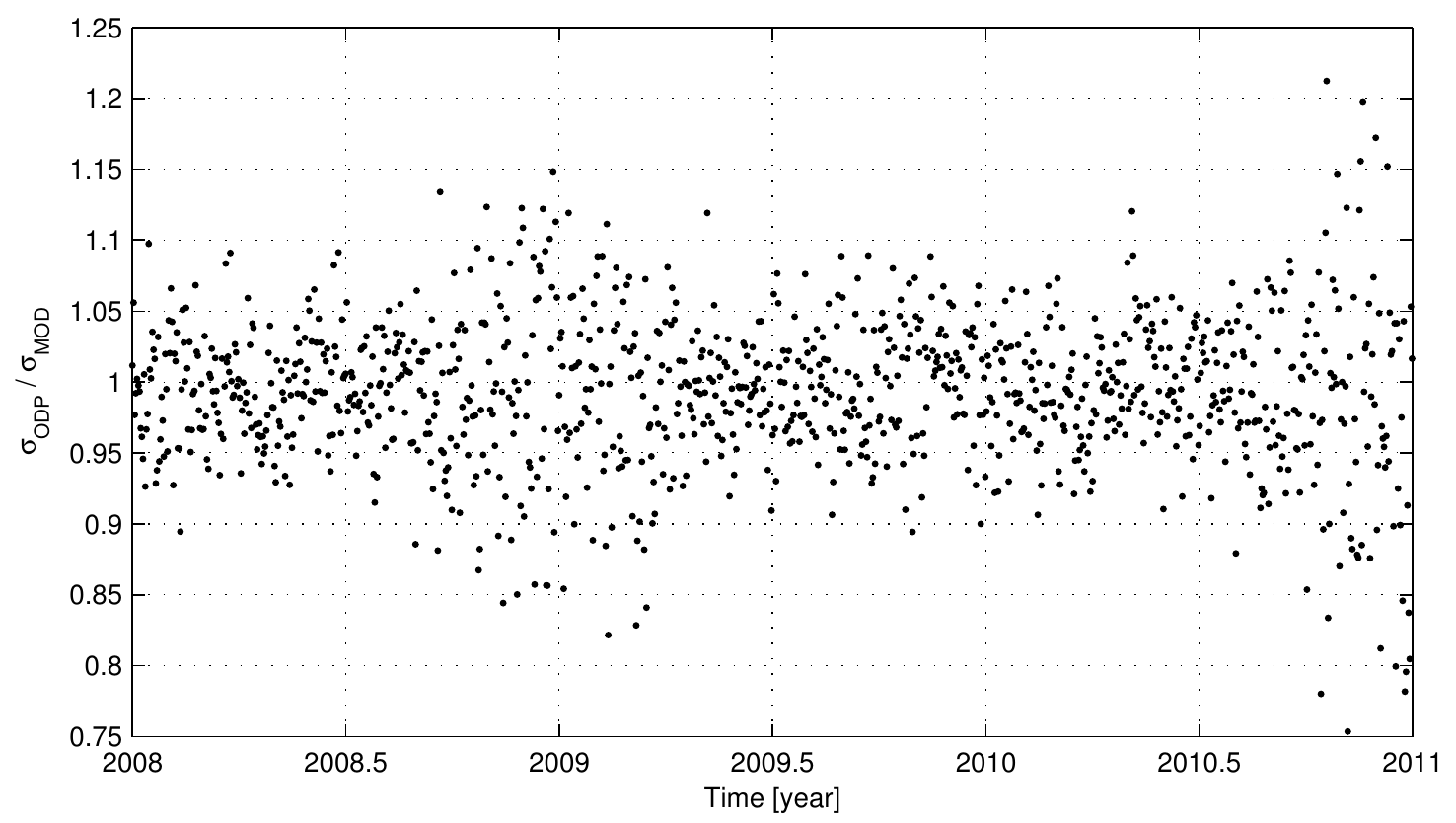}
\caption{\label{fig:sigma_spfodp_mod_rel} Six-parameter fit applied to the Doppler observables computed by the ODP: ratio of the standard deviation of the fit's residuals, $\sigma_{ODP}$, to the expected standard deviation of the numerical errors, $\sigma_{MOD}$, obtained from the final version of the model. Each value is computed on a single tracking pass.}
\end{figure}

\begin{figure}
\centering
\includegraphics[width=\textwidth]{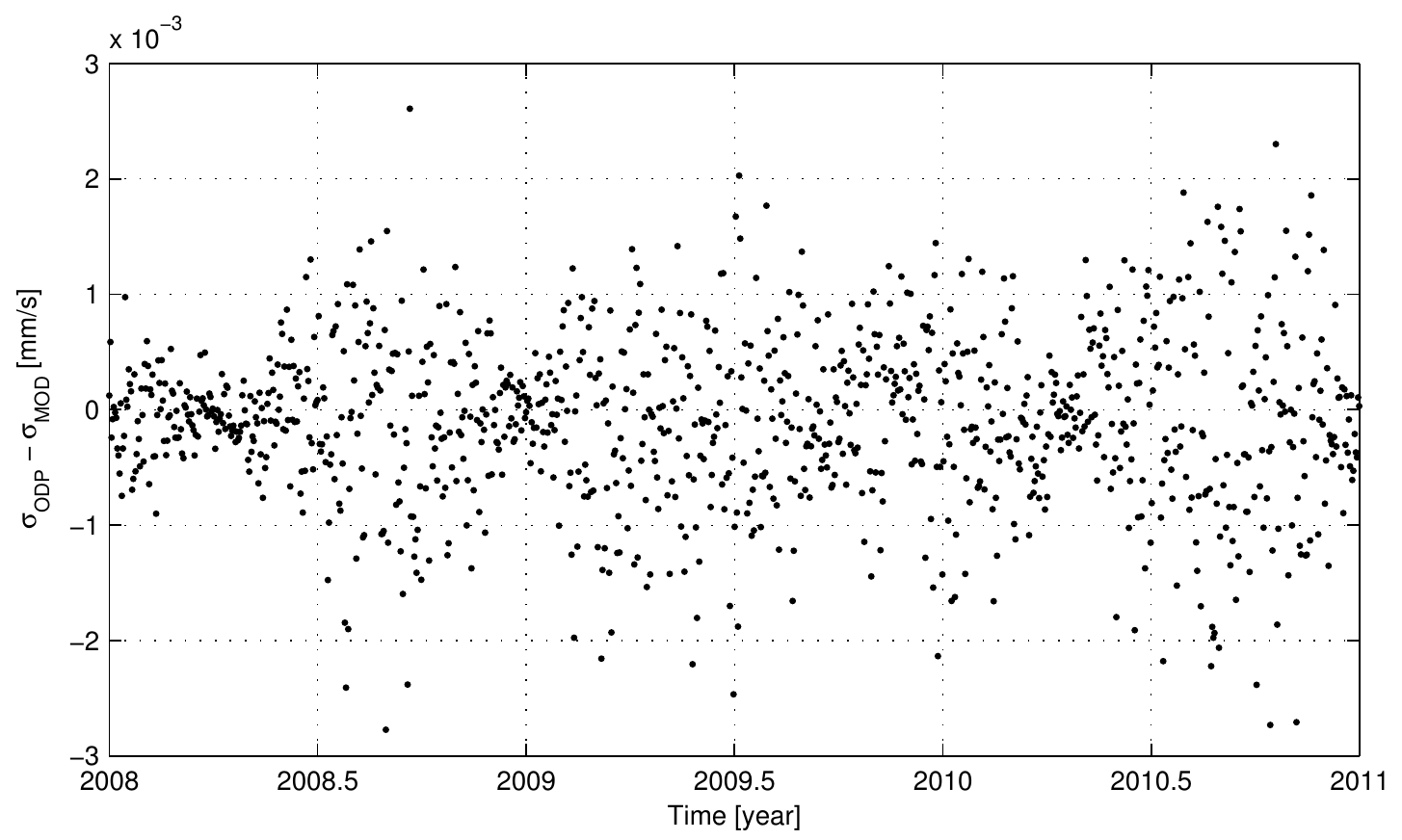}
\caption{\label{fig:sigma_spfodp_mod_diff} Six-parameter fit applied to the Doppler observables computed by the ODP: difference between the standard deviation of the fit's residuals, $\sigma_{ODP}$, and the expected standard deviation of the numerical errors, $\sigma_{MOD}$, obtained from the final version of the model. Each value is computed on a single tracking pass.}
\end{figure}

\begin{figure}
\centering
\includegraphics[width=0.7\textwidth]{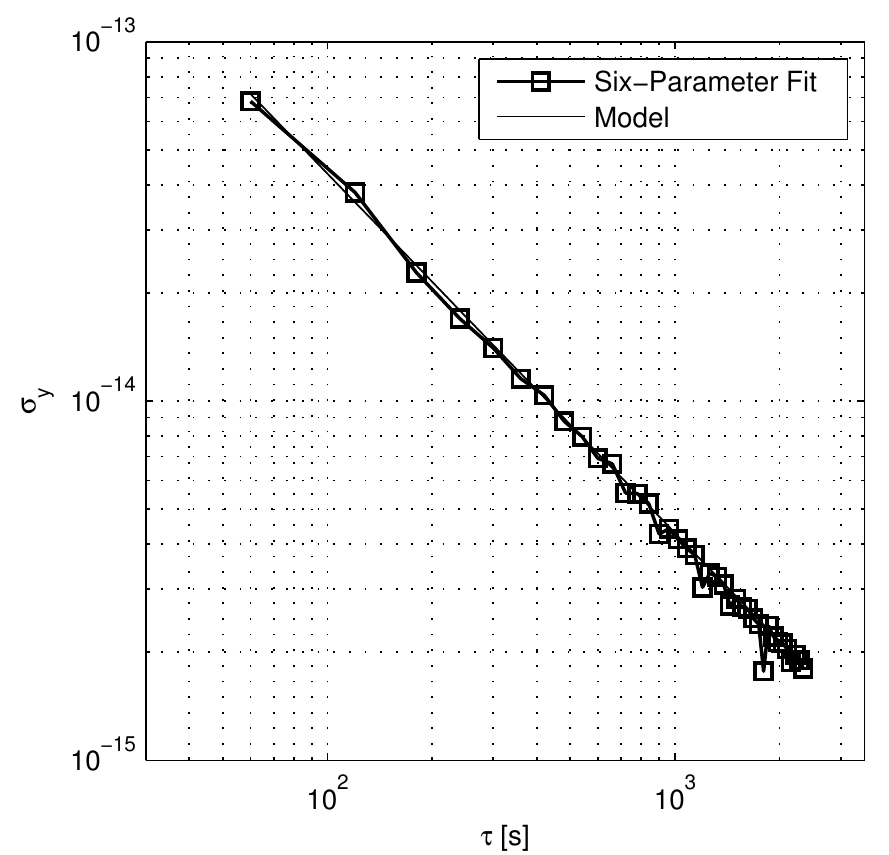}
\caption{\label{fig:asdev_spfodp_mod} Six-parameter fit applied to the Doppler observables computed by the ODP: comparison between the Allan standard deviation of the fit's residuals and the expected Allan standard deviation of the numerical errors, obtained from the final version of the model.}
\end{figure}

For each tracking pass, the numerical noise extracted by the six-parameter fit from the two-way Doppler observables computed by the ODP, was compared to the model, in terms of the following statistical characteristics:
\begin{enumerate}
    \item[1)] Mean: for all analyzed tracking passes the ratio of the residuals' mean to their standard deviation remains bounded within $3\times 10^{-8}$, so it can be considered zero for practical purposes, as predicted by the model.
    \item[2)] Standard Deviation: Fig.~\ref{fig:sigma_spfodp_mod} shows that there is an optimal agreement between the model predictions and the real numerical error. The relative error is almost everywhere below $10\%$, as shown in Fig.~\ref{fig:sigma_spfodp_mod_rel}. The relative difference increases up to $20\%$ when the total numerical errors decrease to small values, so the absolute difference remains globally under $3 \times 10^{-3} \, \text{mm/s}$, as represented in Fig.~\ref{fig:sigma_spfodp_mod_diff}. The increase in the relative difference between the real values and the model is likely due to the approximations and simplifications introduced when developing the prediction model (e.g. the neglected rounding errors and the second order terms in the computation of the partial derivatives).
    \item[3)] ASDEV: it follows very well the expected values and the expected $\tau^{-1}$ slope, as shown in Fig.~\ref{fig:asdev_spfodp_mod}.
\end{enumerate}

Summarizing, the developed model provides quantitative predictions of the different properties of the numerical noise in the Doppler observables. 
These predictions were compared to the actual characteristics of the numerical noise extracted from the ODP. The comparison results, displayed in Figs~\ref{fig:sigma_spfodp_mod}-\ref{fig:asdev_spfodp_mod}, highlight an optimal global agreement between the model and the experimental data.
Moreover, as said in Sec.~\ref{sec:num_err_model}, while developing the model many assumptions were made and different sources of numerical errors were neglected. Given the obtained results, these simplifications were proved not to affect significantly the accuracy of the model predictions.
Hence, the numerical error model can be considered as fully validated at a level of accuracy adequate to \emph{a priori} characterize the expected numerical noise in the radiometric observables of a real OD scenario.

\section{Analysis of Numerical Noise}\label{sec:num_noise_analysis}
\subsection{Case study 1: the Cassini mission}\label{subsec:cassini_num_noise}
The ODP is currently used for navigation and radio science in the Cassini mission to the Saturn system, recently extended to 2017. In particular, in November 2016 Cassini is planned to be inserted into low altitude orbits, characterized by a periapsis of about $1.1$ Saturn radii, just inside the D ring. After about 42 orbits the spacecraft will impact the Saturn atmosphere in September, 2017~\cite{Smith:2009}. Due to the very low periapsis altitude, during this end of mission phase the expected scientific return of Gravity Radio Science Experiments is very relevant.
As the ODP implements only the DRD formulation, the numerical error model described in this paper can be used to assess the expected numerical noise in Cassini's computed Doppler observables.

\begin{figure}
\centering
\includegraphics[width=\textwidth]{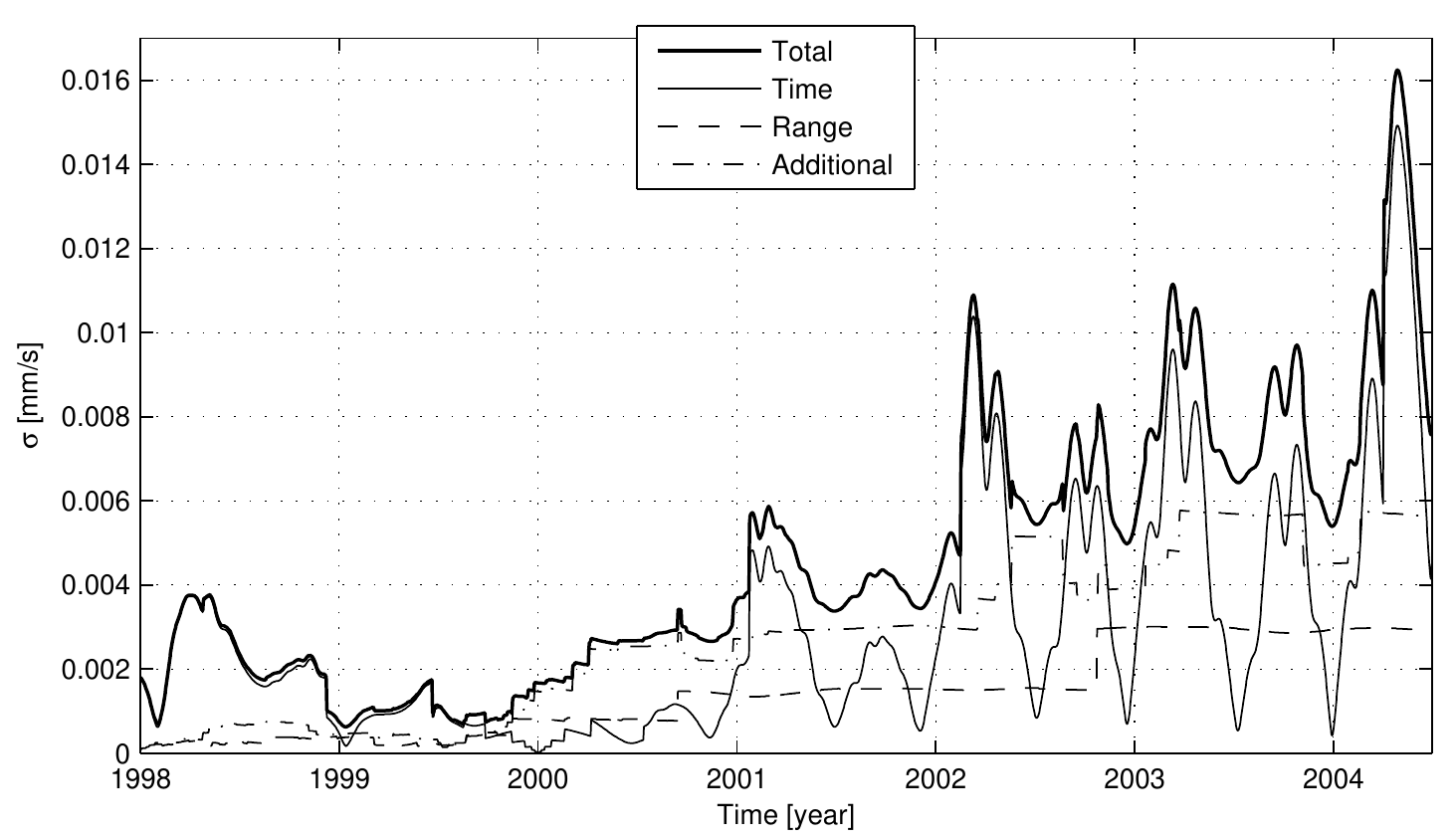}
\caption{\label{fig:cassini_num_noise_cruise} Numerical noise in Cassini's two-way Doppler observables ($T_c$=60\,\text{s}) during the cruise phase, from the beginning of 1998 to the middle of 2004.}
\end{figure}

\begin{figure}
\centering
\includegraphics[width=\textwidth]{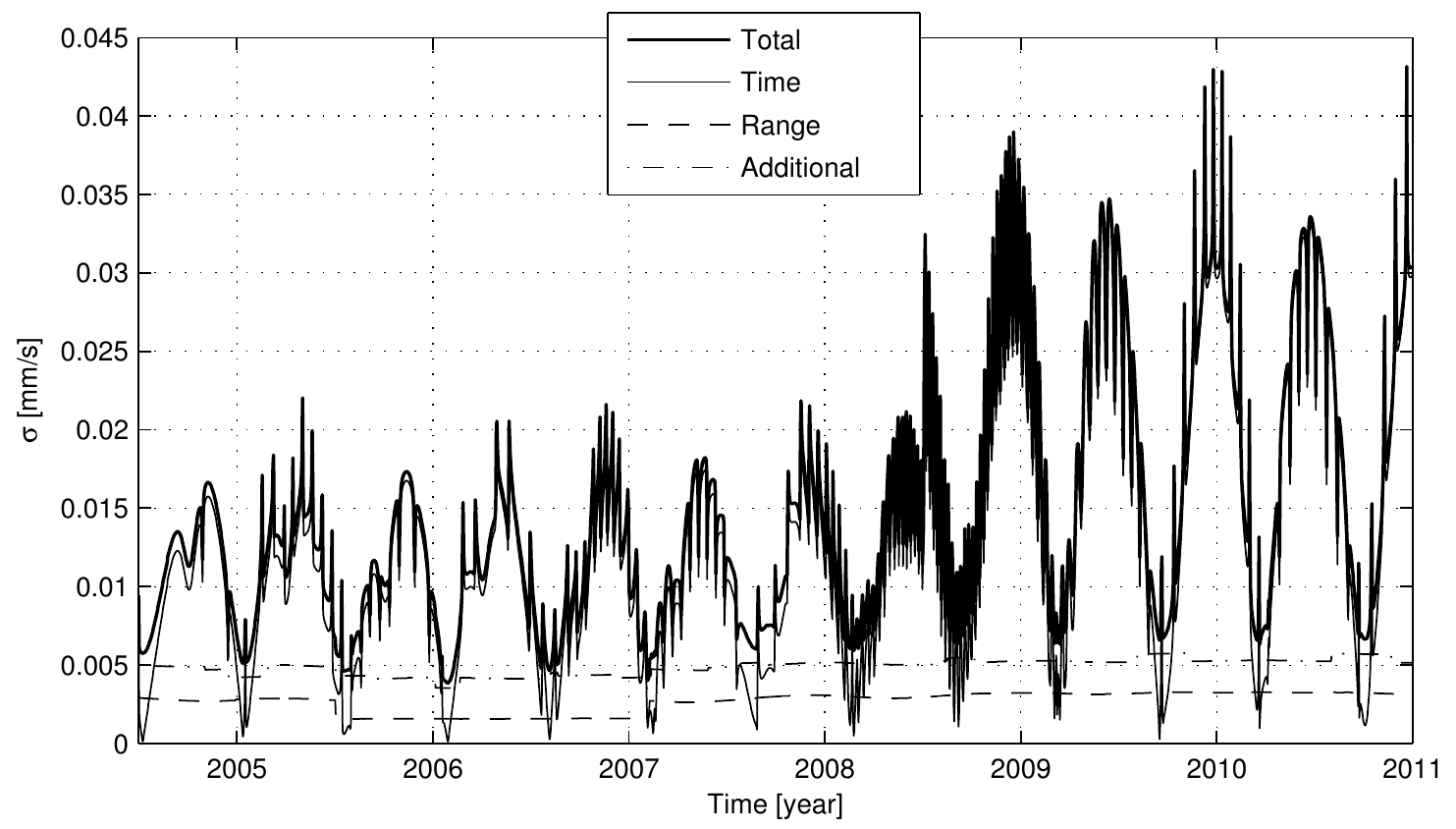}
\caption{\label{fig:cassini_num_noise_nom_ext} Numerical noise in Cassini's two-way Doppler observables ($T_c$=60\,\text{s}) during the Nominal Mission (from the middle of 2004 to the middle of 2008) and the Equinox Mission (from the middle of 2008 to the end of 2010).}
\end{figure}

\begin{figure}
\centering
\includegraphics[width=\textwidth]{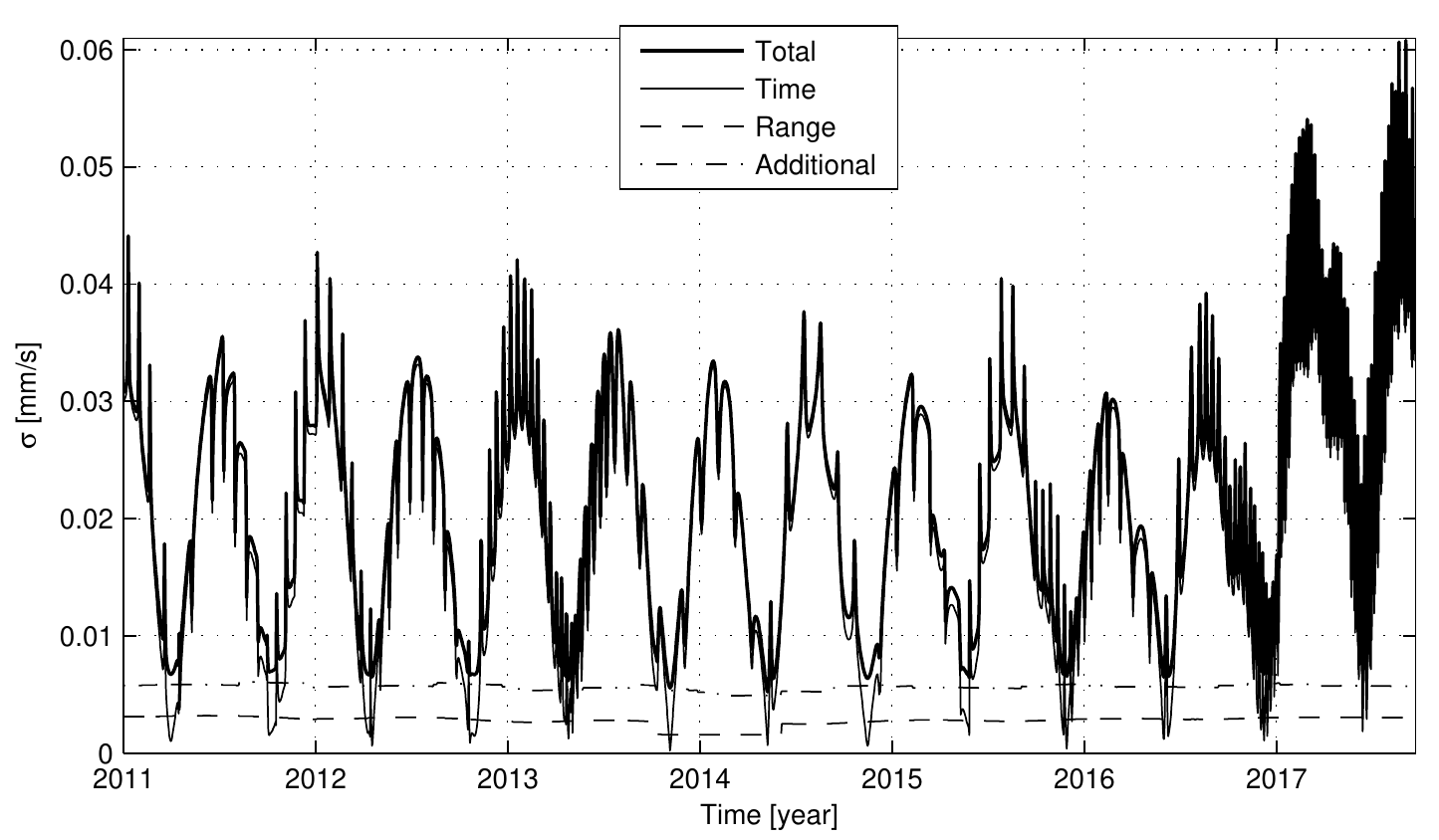}
\caption{\label{fig:cassini_num_noise_extext} Numerical noise in Cassini's two-way Doppler observables ($T_c$=60\,\text{s}) during the Solstice Mission, from the end of 2010 to the end of 2017.}
\end{figure}

The expected numerical noise in Doppler data with a 60-second count time for the entire timespan of the Cassini mission is shown in Fig.~\ref{fig:cassini_num_noise_cruise}, Fig.~\ref{fig:cassini_num_noise_nom_ext}, and Fig.~\ref{fig:cassini_num_noise_extext}. The three plots represent the total numerical noise, plus the noise level due to each of the three error sources: representation of time (Time), position vectors (Range), and additional round-off errors (Additional). Note that the total noise curve is the square root of the sum of the three squared components.
The following comments about Cassini's numerical noise can be made:
\begin{enumerate}
    \item[1)] The Range and Additional components have a similar trend and increase with Cassini's distance from the Earth, reaching almost constant values with the arrival at Saturn's orbit distance.
    \item[2)] The time variation of the Time component is a function of the relative velocity between the ground station and the spacecraft, while the noise amplitude, and the mean level, are a function also of the time elapsed since J2000.
    Hence, the amplitude of the Time component decreases after launch, reaches the minimum value at J2000, and increases from J2000 becoming dominant until the end of mission. In Figs.~\ref{fig:cassini_num_noise_cruise}-\ref{fig:cassini_num_noise_extext} the instantaneous amplitude variation due to the time quantization step changes is clearly visible, for example at the beginning of 2002 and 2004, at the middle of 2008 and at the beginning of 2017.
    During the cruise phase of the Cassini mission, from the beginning of 1998 to the middle of 2004, the trend of the Time component of the numerical noise clearly shows an approximately half-year periodicity due to Earth's orbital motion. After the Saturn Orbit Insertion (SOI), at the middle of 2004, the trend of the Time component is the superposition of the low-frequency Earth-Saturn relative motion and the higher frequency Cassini orbital motion around Saturn. As the figures were generated creating a single data point per day, the effect of the Earth's rotation, which has a maximum value of about $4 \times 10^{-5} \,\text{mm/s}$, is filtered out.
    \item[3)] From the comparison of the three numerical error components, it can be stated that during the cruise phase, until the beginning of 2004, the three components had, on average, about the same order of magnitude. Since about the beginning of 2004, the Time component has increased and became dominant over the other sources, in almost the entire residual mission.
    \item[4)] The current best accuracy in Cassini two-way Doppler tracking, with proper state-of-the-art calibrations and under favorable conditions, is about $9 \times 10^{-4}\,\text{mm/s}$ with a count time of $1000 \,\text{s}$, and it was achieved during the Gravitational Wave Experiment (GWE), performed during the 2001-2002 solar opposition~\cite{Armstrong:2003}. As another reference, during the Solar Conjunction Experiment (SCE) at the middle of 2002, a two-way fractional frequency stability of about $4.3\times 10^{-3}\,\text{mm/s}$ was achieved, with a count time of $300 \,\text{s}$~\cite{Bertotti:2003}.
    Assuming white phase noise, the corresponding noise level at $60 \, \text{s}$ integration time is about $3.7 \times 10^{-3}\,\text{mm/s}$ for the GWE and about $9.5 \times 10^{-3} \, \text{mm/s}$ for the SCE.
    The expected numerical noise, with the same count time, has been of the same order of magnitude since about the beginning of 2001, and becomes up to 15 times larger at the beginning of 2017, near the end of mission.
    Note that this comparison does not reflect the numerical noise level experienced during these experiments, because they were performed using higher count times. White phase noise scales with $T_c^{-\frac{1}{2}}$, while the numerical noise scales with $T_c^{-1}$. For example, during GWE an accuracy of about $9 \times 10^{-4}\,\text{mm/s}$ was achieved, with a count time of $1000 \,\text{s}$. At this count time the corresponding expected numerical noise is $~ 4 \times 10^{-3} \cdot 60/1000 = 2.4 \times 10^{-4} \, \text{mm/s}$.
\end{enumerate}
Summarizing, the numerical noise is a very important source of error in Cassini's OD carried out using the ODP, and will become a critical factor in 2017, during the proximal orbits phase.

\subsection{Case study 2: the Juno mission}\label{subsec:juno_num_noise}
Juno is a NASA New Frontiers interplanetary mission to study the origin, interior structure, and evolution of the planet Jupiter~\cite{Matousek:2007}.
Juno was launched on August 15, 2011, and, after two Deep Space Maneuvers (DSM) and an Earth gravity assist, will reach Jupiter in July 2016. 
In August 2016, through the Jupiter Orbit Insertion (JOI) maneuver, it will be placed in a highly eccentric polar orbit around Jupiter.
The nominal mission will end in September 2017, with an impact on Jupiter after a de-orbiting maneuver.
The mission will consist of 31 science orbits, 25 of which will be dedicated to gravity radio science experiments.
In fact, one of the main scientific objectives of the Juno mission is to map the Jupiter gravity field with unprecedented accuracy. During gravity orbits, a dual-frequency (at X- and Ka-band) two-way Doppler link will be established between the Goldstone Deep Space Network (DSN) complex\footnote{At present, only the Deep Space Station (DSS) 25 of the Goldstone site is capable of transmitting at Ka-band.} and the spacecraft Juno, with a nominal observing interval of about 6 hours around perijove. This configuration, along with state-of-the-art troposphere calibration techniques, is expected to reach a two-way accuracy of $1.0 \times 10^{-2} \, \text{mm/s}$ at $60 \, \text{s}$ integration time, during the entire mission~\cite{Anderson:2004}.

As in the Cassini case study, the developed numerical error model can be used to assess the expected numerical noise in Juno's computed two-way Doppler observables.

\begin{figure}
\centering
\includegraphics[width=\textwidth]{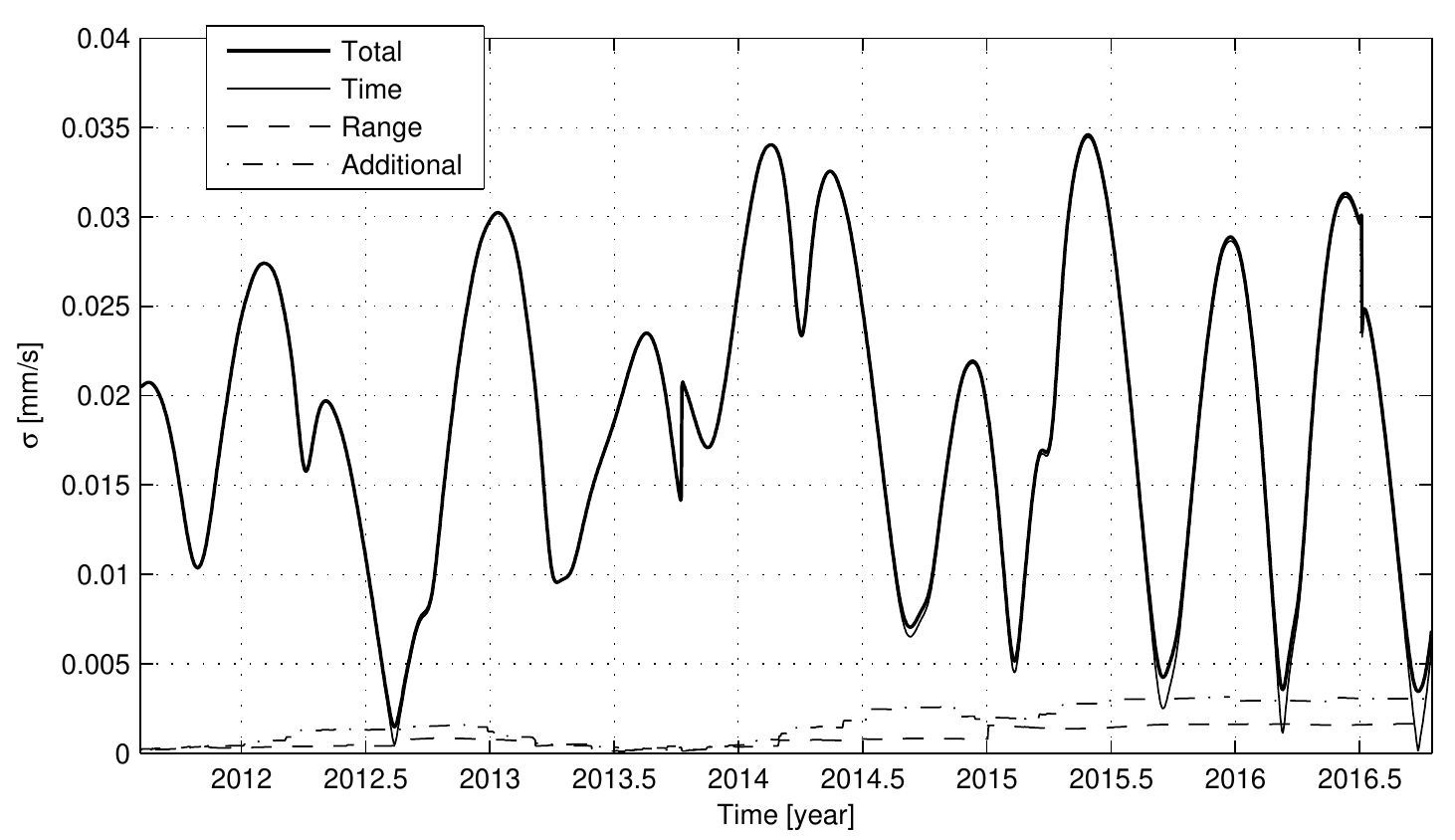}
\caption{\label{fig:juno_num_noise_60_cruise} Numerical noise in Juno's two-way Doppler observables ($T_c$=60\,\text{s}) during the cruise phase, from the end of 2011 to the end of 2016.}
\end{figure}

\begin{figure}
\centering
\includegraphics[width=\textwidth]{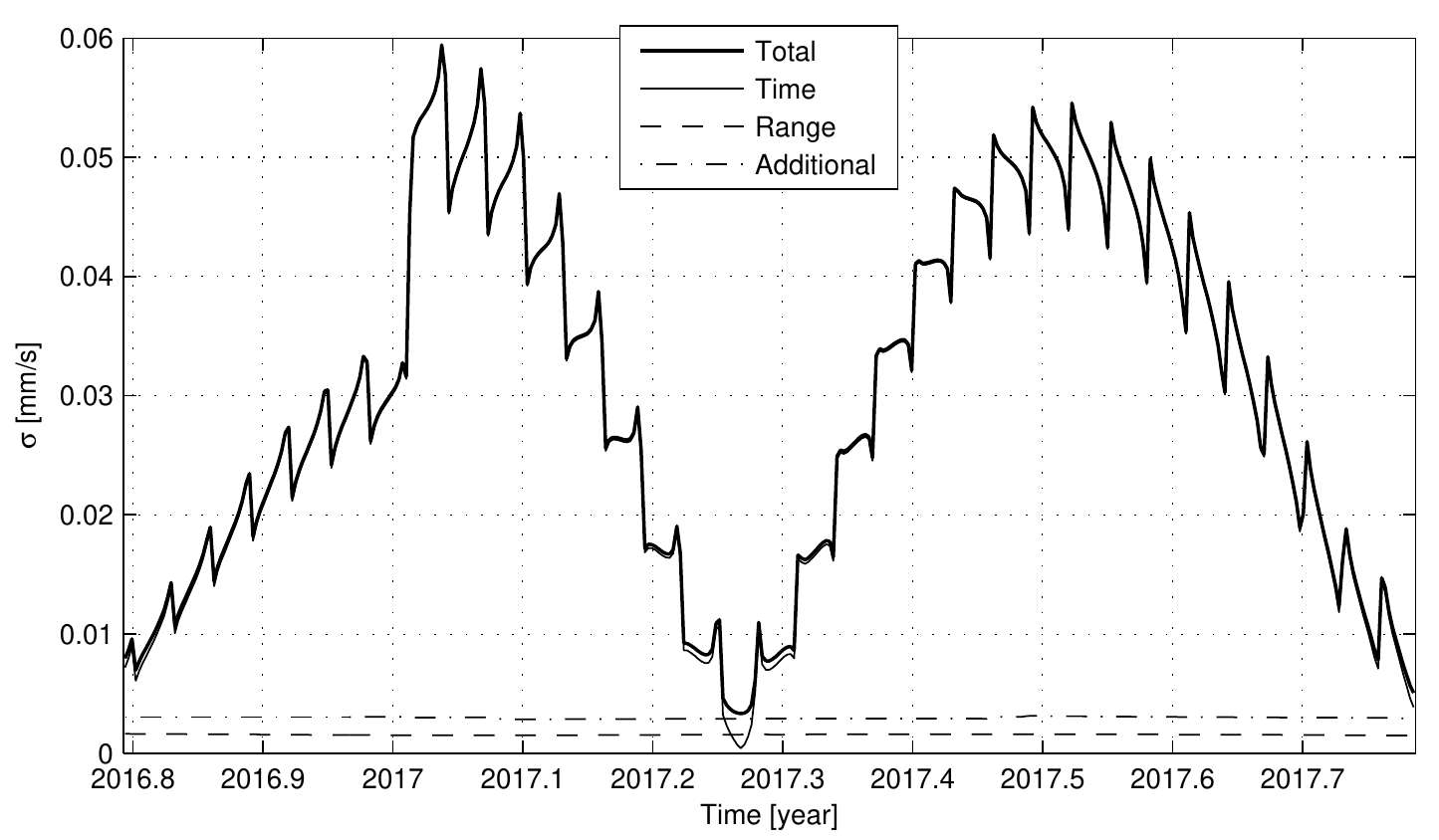}
\caption{\label{fig:juno_num_noise_60_nom} Numerical noise in Juno's two-way Doppler observables ($T_c$=60\,\text{s}) during the Nominal Mission (from the end of 2016 to the end of 2017).}
\end{figure}

Fig.~\ref{fig:juno_num_noise_60_cruise} and Fig.~\ref{fig:juno_num_noise_60_nom} show the expected numerical noise in two-way Doppler data with a count time of $60 \, \text{s}$, during the interplanetary cruise and nominal mission phases, respectively.
As for the Cassini case study, the plots represent the total numerical noise and the three sources of numerical errors: representation of time (Time), position vectors (Range), and additional round-off errors (Additional).

For the Juno mission, considerations similar to those for the Cassini mission apply. In summary:
\begin{enumerate}
    \item[1)] The numerical noise is dominated by the Time component during almost the entire mission duration. Comparing the Juno mission to the Cassini Solstice mission, the time variable assumes approximately the same values, but in the Juno mission the Range and Additional components have less influence on the total noise because of the shorter distance from the SSB (about $5.4 \, \text{AU}$ for SSB-Jupiter distance, about $10 \, \text{AU}$ for SSB-Saturn distance).
    \item[2)] The Time component reflects the range-rate between the Earth and the spacecraft. In particular, during the nominal mission phase, the total range-rate is the superposition of the Earth-Jupiter relative motion, characterized by a period of about half a year, and the Jupiter-Juno orbital motion, characterized by a period of about $11 \, \text{days}$. As expected, at the beginning of 2017 the Time component instantaneously increases, and its value approximately doubles.
    \item[3)] For a large portion of the mission, the total numerical noise is significantly larger than the expected two-way Doppler measurement accuracy of $1.0 \times 10^{-2} \, \text{mm/s}$ at $60 \, \text{s}$ integration time. Hence, in order to fully exploit the capabilities of the high performance Juno Doppler link, it is necessary to reduce the numerical noise impact in the Doppler observables computed by the OD program.
\end{enumerate}

\section{Numerical Noise Reduction Strategies}\label{sec:num_noise_reduction}
On the basis of the developed model, three main strategies were identified to mitigate the numerical errors in the generation of the computed values of radiometric observables:
\begin{enumerate}
    \item[1)] Compile the OD program using a higher precision floating point representation. Moving from the double-precision to quadruple-precision representation, all round-off errors decrease by about a factor of $10^{18}$, becoming completely negligible. As a side effect, the execution time of the OD process increases. However, because of the high complexity of the OD codes, the increase in the execution time cannot be evaluated \emph{a priori}, but must be assessed through dedicated experimental tests. Several approaches to reduce the execution time can be applied, like parallelization or recompilation in quadruple-precision of a limited subset of procedures, with further effects of increasing the complexity of the source code.
    \item[2)] Change the representation of time. As shown in Section~\ref{sec:num_noise_analysis} for the Cassini and Juno missions, the Time component of numerical noise represents the most important contribution in interplanetary OD problems. Hence, the numerical noise can be significantly reduced by changing only the time representation. Currently, both the ODP and AMFIN use a single double-precision variable to represent time. The accuracy in time representation could be increased using two variables in pairs, $(n,t_m)$: an integer $n$, representing the number of days elapsed since a reference epoch (such as J2000), and a double-precision value $t_m$, representing the time, in a defined measurement unit, elapsed since 00:00 of the current day.
    The next generation NASA/JPL's OD program, ``Mission-analysis, Operations, and Navigation Toolkit Environment'' (MONTE), is also based upon the Moyer DRD formulation, but makes use of this new ``extended-precision'' time representation.
    As $t_m \le 86400 \,\text{s}$, the maximum time quantization step with MONTE will be about $1.5 \times 10^{-11} \,\text{s}$, about $8200$ times smaller than the maximum quantization step of a double-precision time variable, measured in seconds past J2000, between 2017 and 2034. The corresponding effect in computed values of the radiometric observables is perfectly negligible. However, the round-off errors in the other numerical noise components (Additional and Range) remain unchanged, so they become the most important numerical error sources. The resulting numerical noise is roughly a function of the distance between the spacecraft and the tracking station. For example, for Cassini two-way Doppler data with $T_c = 60 \,\text{s}$, the maximum numerical noise will decrease from about $6 \times 10^{-2} \,\text{mm/s}$, reached at end-of-mission, to about $6 \times 10^{-3} \,\text{mm/s}$, nearly constant since Saturn Orbit Insertion at the middle of 2004. Similarly, for Juno two-way Doppler data with $T_c = 60 \, \text{s}$, the maximum numerical noise will decrease from about $6 \times 10^{-2} \, \text{mm/s}$, computed at end of mission, to about $3.5 \times 10^{-3} \, \text{mm/s}$, constant from the middle of 2015. In both cases the numerical noise will become smaller than the best expected Doppler measurement precision, still remaining a non-negligible error source. As a side effect, each operation explicitly involving time variables must be performed using custom functions, increasing the complexity of the source code. For example, the time interval between two epochs could not be computed as a simple difference, but must be performed through a dedicated function.
    \item[3)] Use the Integrated Doppler formulation, which is much less influenced by numerical errors~\cite{Moyer:1969}. According to this formulation, the Doppler observable is computed expanding the Doppler frequency shift in a Taylor time series. Averaging over the count time, terms proportional to the odd powers of $T_c$ become zero. Hence, retaining only terms up to $T_c^2$, the intrinsic error of the ID formulation increases with $T_c^4$. The maximum allowable count time depends on the specific OD problem: for a desired accuracy of $10^{-2} \, \text{mm/s}$, the count time has to be less than 1-10\,s if the spacecraft is close to a planet or a satellite, while in heliocentric cruise much longer count times can be used, up to $1000 \, \text{s}$~\cite{Moyer:1969}. On the contrary, the DRD formulation is theoretically exact, but is very sensitive to numerical errors, which increase as $T_c^{-1}$ when reducing the count time. Hence, the numerical noise could be reduced using the ID formulation, but only at small count times. Moreover, the ID formulation can be only used with a constant uplink frequency. For most modern and future interplanetary missions, the uplink carrier frequency is varied in order to minimize the Doppler frequency shift observed by the spacecraft. The ID formulation is currently implemented only in AMFIN. It was implemented in older versions of the ODP, but it was then replaced by the DRD formulation.
\end{enumerate}

\section{Conclusions}\label{sec:conclusions}
In this paper a model of numerical errors in two-way and three-way Doppler observables computed using the Moyer differenced-range Doppler (DRD) formulation was described. The DRD formultation is currently implemented in the most important interplanetary orbit determination (OD) program: NASA/JPL's ``Orbit Determination Program'' (ODP) and ``Mission-analysis, Operations, and Navigation Toolkit Environment'' (MONTE), and ESA's ``Advanced Modular Facility for Interplanetary Navigation'' (AMFIN). The model was validated analyzing directly the numerical noise in Doppler observables computed by the ODP, extracted using a simplified fitting function, which was referred to as the ``six-parameter fit''. In particular, the model showed an accuracy always better than $3 \times 10^{-3}\,\text{mm/s}$ in the estimation of the numerical noise standard deviation, at $60 \,\text{s}$ integration time.

An accurate prediction of numerical noise can be used to compute a proper noise budget in Doppler tracking of interplanetary spacecraft. This represents a critical step for the design of future interplanetary missions, both for radio science experiments, requiring the highest level of accuracy, and spacecraft navigation. Moreover, the accurate prediction of numerical noise can also be used to identify enhancements in past radio science experiments, if an improved OD code, less affected by numerical errors, could be used to reprocess past archived data.
On the basis of the numerical errors characterization, three different approaches to reduce the numerical noise were proposed.

As real-world scenario case studies, the expected numerical noise in the two-way Doppler link of the Cassini and Juno interplanetary missions was analyzed. As a result, the numerical noise proved to be, in general, not negligible. Furthermore, in some conditions, the numerical noise can be the dominant noise source in the OD process. Hence, the introduction of a reduced-numerical-noise OD program is considered mandatory, not only for future interplanetary missions to the outer planets, but also for the currently operational Cassini mission.

\section*{Acknowledgments}
The authors would like to thank Frank Budnik from ESA/ESOC's Flight Dynamics team, Marco Lanucara and Mattia Mercolino from ESA/ESOC's Systems and Project Support Section, and Luciano Iess from ``Sapienza'' University of Rome, for the useful discussions on Doppler tracking error budgets. We are also grateful to John W. Armstrong and Frederic J. Pelletier from Jet Propulsion Laboratory, California Institute of Technology, for their careful revision of the manuscript. This work was funded in part by the Italian Space Agency (ASI) through Cassini-Huygens contract AMM-IAF-RM 02/2010.

\end{document}